\documentclass[a4paper,11pt,twoside]{article}

\usepackage{amsmath}
\usepackage{amssymb}
\usepackage{epsfig}

\newcommand{\interskip}{\medskip}

\newcommand{\MeV}{\,\mathrm{MeV}}

\newcommand{\eV}{\,\mathrm{eV}}

\newcommand{\ord}[1]{\mathcal{O}\left( #1 \right)}

\DeclareMathOperator{\diag}{diag}

\newcommand{\dm}[1]{{\Delta m^2_{\text{#1}}}}

\newcommand{\capdef}{}
\newcommand{\mycaption}[2][\capdef]{\renewcommand{\capdef}{#2}%
        \caption[#1]{{\itshape #2}}} 
\makeatletter
\renewcommand{\fnum@table}{\textbf{\tablename~\thetable}}
\renewcommand{\fnum@figure}{\textbf{\figurename~\thefigure}}
\makeatother

\newlength{\myem}
\newcommand{\sep}[1]{#1}
\newcounter{mysubequation}[equation]
\renewcommand{\themysubequation}{\alph{mysubequation}}
\newcommand{\mytag}{\stepcounter{mysubequation}%
\tag{\theequation\protect\sep{\themysubequation}}}
\newcommand{\globallabel}[1]{\refstepcounter{equation}\label{#1}}

\makeatletter
\renewcommand{\section}{\@startsection{section}{1}{0em}%
        {-3.5ex \@plus -1ex \@minus -.2ex}% 
        {2.3ex \@plus.2ex}%
        {\normalfont\large\bfseries}}
\renewcommand{\subsection}{\@startsection{subsection}{2}{0em}%
        {-3.25ex\@plus -1ex \@minus -.2ex}%
        {1.5ex \@plus .2ex}%
        {\normalfont\bfseries}}
\renewcommand{\subsubsection}%
        {\@startsection{subsubsection}{3}{0em}%
        {-3.25ex\@plus -1ex \@minus -.2ex}%
        {1.5ex \@plus .2ex}%
        {\normalfont\itshape}}
\makeatother

\newcommand{\Z}{\mathbb{Z}}
\newcommand{\ZZ}{\mathbb{Z}_2}

\newcommand{\Fig}[1]{Fig.~\ref{fig:#1}}

\newcommand{\Eq}[1]{Eq.~(\ref{#1})}
\newcommand{\eq}[1]{eq.~(\ref{#1})}
\newcommand{\Eqs}[1]{Eqs.~(\ref{#1})}
\newcommand{\eqs}[1]{eqs.~(\ref{#1})}

\newcommand{\dash}{\,\text{--}\,}
\newlength{\phantomlength}
\newcommand{\phantomheigth}[1]{%
  \settowidth{\phantomlength}{$\displaystyle #1$}%
  \phantom{#1}\hspace*{-\phantomlength}}
\newcommand{\phantombox}[2]{%
  \settowidth{\phantomlength}{#1}%
  \makebox[\phantomlength]{#2}}
\newcommand{\ml}{{m'}}
\newcommand{\mbb}{m}
\newcommand{\mbbc}{m^c}
\newcommand{\MM}{M}
\newcommand{\MD}{\mu}
\newcommand{\nuh}{N}
\newcommand{\bv}{\mathbf{v}}
\newcommand{\bD}{\mathbf{D}}
\newcommand{\bnu}{\mathbf{\nu}}
\newcommand{\bchi}{\mathbf{\chi}}
\newcommand{\balpha}{\mathbf{\alpha}}
\newcommand{\bbeta}{\mathbf{\beta}}

\newcommand{\bea}{\begin{eqnarray}}
\newcommand{\eea}{\end{eqnarray}}

\def\a{\alpha}
\def\b{\beta}
\def\g{\gamma}

\def\d{\delta}
\def\e{\epsilon}

\def\l{\lambda}
\def\m{\mu}
\def\n{\nu}
\def\o{\omega}
\def\p{\pi}

\def\r{\rho}
\def\s{\sigma}

\def\t{\tau}

\def\x{\xi}

\def\D{\Delta}

\def\cL{{\cal L}}
\def\cM{{\cal M}}
\def\cN{{\cal N}}
\def\um{\mathbf{1}}

\makeatletter
\@addtoreset{equation}{section}
\makeatother
\renewcommand{\theequation}{\thesection.\arabic{equation}}

\newcommand{\OX}{Department of Physics, Theoretical Physics,
University of Oxford, Oxford OX1\hspace{0.2em}3NP, UK}
\newcommand{\FL}{Institute of Fundamental Theory, Department of
Physics\\ University of Florida, Gainesville, FL 32611, USA}

\newcommand{\FNAL}{Fermi National Accelerator Laboratory\\ P.O. Box 500,
Batavia, IL 60510, USA}
\newcommand{\Sussex}{Centre for Theoretical Physics, University of Sussex\\
Falmer, Brighton BN1 9QJ, UK}

\newcommand{\preprintdate}{November 2000}
\newcommand{\preprintnumber}{FERMILAB-Pub-00/284-T\\OUTP-00-39P\\
            SUSX-TH/00-018\\UFIFT-HET00-27}
 
\newcommand{\titletext}{Neutrino Masses and Mixing in Brane-World Theories} 
\newcommand{\authortext}{\large Andr\'e Lukas~$^\star$, Pierre
                         Ramond~$^\#$, Andrea Romanino~$^\bullet$ and
                         Graham G. Ross~$^\diamond$\medskip\\\em\normalsize
                         $^\star$~\Sussex\medskip\\$^\#$~\FL\medskip\\
                         $^\bullet$~\FNAL\medskip\\$^\diamond$~\OX}
\newcommand{\abstracttext}{
We present a comprehensive study of five-dimensional brane-world
models for neutrino physics based on flat compactifications.
Particular emphasis is put on the inclusion of bulk mass terms.
We derive a number of general results for such brane-world models
with bulk mass terms. In particular, in the limit of
small brane-bulk couplings, the electroweak eigenstates are predominantly
given as a superposition of three light states with non-trivial
small admixtures of bulk states. As a consequence, neutrinos can
undergo standard oscillations as well as oscillation into bulk
Kaluza-Klein states. We use this structure to construct a specific
model based on $\ZZ$ orbifolding and bulk Majorana masses which is
compatible with all observed oscillation phenomena. The solar
neutrino deficit is explained by oscillations into sterile bulk
states while the atmospheric neutrino deficit is due to $\nu_\m$
-- $\nu_\tau$ oscillations with naturally maximal mixing. In addition,
the model can accommodate the LSND result and a significant
neutrino dark matter component. We also analyze the constraints from
supernova energy loss on neutrino brane-world theories and show that
our specific model is consistent with these constraints.}

\title{
\normalsize
\begin{tabular}[t]{l}\preprintdate\end{tabular}
\hspace*{\fill}
\begin{tabular}[t]{l}\preprintnumber\end{tabular}
\vspace{3\baselineskip}\\\Large\bfseries\titletext\bigskip}
\author{\begin{minipage}[t]{0.8\textwidth}
\normalsize\centering\authortext
\end{minipage}}
\date{}

\begin{document}

\bigskip
\maketitle
\begin{abstract}\normalsize\noindent\abstracttext\end{abstract}
\normalsize\vspace{\baselineskip}
\thispagestyle{empty}
\clearpage

\setcounter{page}{1}

\section{Introduction}
\label{sec:introduction}

\noindent

Two recent ideas, namely the brane-world idea~\cite{Horava:96a,
Witten:96a,Horava:96b,Lukas:98b,Lukas:98c,Arkani-Hamed:98a,Antoniadis:98a,
Kakushadze:98a} and the
possibility of having large gravitation-only additional
dimensions~\cite{Witten:96a,Lykken:96a,Arkani-Hamed:98a,Antoniadis:98a},
may significantly change the relation
between string- or M-theory and low-energy particle physics.
Not only do these ideas motivate new directions in particle physics
model building but they also provide new generic structures within
string theory which may be experimentally testable. In this context,
an important role is played by bulk particles coupling to standard
model particles only gravitationally. Particularly relevant are
the Kaluza-Klein modes of the higher-dimensional
graviton~\cite{Arkani-Hamed:98a} and higher-dimensional bulk fermions.
In this paper, we focus on the latter possibility of bulk fermions.
They provide candidates for right-handed neutrinos and might,
therefore, play an important role in neutrinos physics.

The potential relevance of these particles for neutrino physics has
first been first pointed out and analyzed in
Ref.~\cite{Dienes:98a,Arkani-Hamed:98c,Dvali:99a}. Phenomenological
analysis have been performed
in~\cite{Faraggi:1999bm}--\cite{Abazajian:2000hw}.
The importance of bulk masses has been pointed out in~\cite{LR}, where
an explicit 6-dimensional example with Dirac bulk masses has been
presented. The phenomenological implications of a five-dimensional
brane-world model with bulk Dirac-masses for solar, atmospheric,
short-baseline and supernova oscillations have been studied
in~\cite{LRRR}.

The main goal of this paper is to present a comprehensive
phenomenological study of five-dimensional brane-world models for
neutrino mass and mixing. Particular emphasis will be put on the
model-building options that arise from the various types of bulk mass
terms.  We will focus on compactifications on a flat metric as well as
on bulk mass terms without any explicit dependence on the additional
coordinate.  Furthermore, we assume that all right-handed neutrinos
propagate in the fifth dimension.  Within this general setting we will
present a number of exact results for the mass spectrum as well as the
mixing angles.  Furthermore, based on these general results, we will
give an explicit example based on the orbifold $S^1/\ZZ$ and Majorana
bulk-masses related to the $S^1$ model with Dirac masses studied in
Ref.~\cite{LRRR}.  As we will see, this model is consistent with the
oscillation data presently available. In particular, we show that the
solar neutrino deficit can be explained by small-mixing angle
oscillations of $\nu_e$ into a tower of sterile neutrinos. Unlike such
oscillations into a single sterile neutrino this option is not
disfavored by the recent SuperKamiokande results. In addition, the
model naturally leads to a maximal mixing angle in the $\n_\m$ --
$\n_\tau$ sector and can explain the atmospheric neutrino results. At
the same time it contains two neutrinos in the eV range that can
provide a cosmologically significant source of dark matter. It also
has room to accommodate the LSND signal and in general predicts
$\nu_e\leftrightarrow\nu_\mu$ oscillations with $\dm{}> \dm{ATM}$ and
a small amplitude. The alternative model-building option, namely large
mixing angles between standard and sterile Kaluza-Klein neutrinos,
is disfavored by the requirement of not having significant departures
from Standard Model processes~\cite{ioannisian}. In particular, this
disfavors the possibility that sterile bulk states play a significant
role in atmosperic neutrino oscillations.

\medskip

In the context of string- and M-theory, bulk fermions arise as
superpartners of gravitational moduli, such as, for example, radii of
internal spaces. Given this origin, the existence of bulk
fermions is practically unavoidable in any supersymmetric string
compactification and, therefore, represents a quite generic feature of
string theory. This, in our opinion, constitutes the most likely
origin of such particles within a fundamental theory and, at the same time,
provides the best theoretical motivation to study brane-world
neutrino physics. In the context of four-dimensional effective
actions from string theory, modulini as candidates for right-handed
neutrinos have been proposed in Ref.~\cite{BS}.
The phenomenological relevance of such bulk fermions for neutrino
physics is basically controlled by three different features of the
theory. Specifically, these are the masses of the bulk fermions, the
radii of the additional dimensions and the size of the brane-bulk
coupling between bulk fermions and standard model neutrinos. Let us
discuss these three features and their relation to string- and
M-theory separately.

So far, brane-world model building in the context of neutrino physics
has been mostly focusing on exactly massless bulk fermions (see,
however, Ref.~\cite{Dienes:98a,Ioannisian:99a} where bulk mass
terms have been considered). However, bulk mass terms constitute an important
model-building option which should be taken into consideration. In
fact, unless forbidden by specific symmetries, bulk mass terms
should not be ignored. Of course, bulk fermions can be candidates for
sterile neutrinos only if these masses are sufficiently
small~\footnote{As we will see in Section 2, only the smallness of 
Lorentz-invariant mass terms needs to be accounted for.}.
One should, therefore, worry about the origin of such small mass scales.
Within perturbative string theory, moduli fields constitute flat
directions and, consequently, have vanishing mass. The same is true
for their fermionic superpartners as long as supersymmetry is
unbroken. String theory, therefore, provides a generic reason why
bulk fermion masses may be much smaller than the string scale.
Eventually, non-perturbative effects have to be taken into account
in order to break supersymmetry and stabilize the moduli fields.
Those will generate masses for the moduli field and, most likely,
masses for the bulk fermions as well. Hence, the inclusion of small
bulk fermion masses is quite well motivated from the viewpoint of
string theory. It is one of the central points of this paper, to
take this insight seriously and include the most general structure
of bulk mass terms into our analysis.

The specific size of these mass terms as well as the size of the
additional dimension depend on non-perturbative effects and are
largely uncertain from a theoretical perspective. Therefore, in the
general part of the paper, we will not focus on any specific choice
for these scales.  Of course, the tower of bulk fermions will only be
directly relevant for oscillations if the bulk masses and the scale of
the additional dimension are sufficiently small. For the explicit
example we will focus on such a case with small bulk masses and a
large fifth dimension close to the experimental upper bound.

Finally, we should address the possible origin of the brane-bulk mass
terms mixing bulk fermions and standard neutrinos.  In fact, such
couplings may arise quite generically, for example, in the context of
Horava-Witten theory~\cite{Horava:96a}. In the 11-dimensional
formulation of this theory, the 10-dimensional Yang-Mills theories on
the boundaries contain terms bilinear in the gravitino (giving rise to
bulk fermions in lower dimensions) and the gauginos (giving rise to
the standard model fermions). These are precisely the terms needed to
generate the desired brane-bulk mixing. From the perspective of
effective four-dimensional supergravity theories such terms can be
understood as arising from higher-dimensional non-renormalizable
operators involving moduli and standard model fields with appropriate
VEVs inserted.  It would certainly be interesting to analyze this in
more detail.  However, for the purpose of the present paper, we
content ourselves with these general remarks.

We would like to point out that there exist models, particularly in the
context of string compactifications, where, in addition to bulk
right-handed neutrinos, one also has right-handed neutrinos on the
brane. Such models are problematic for values of the fundamental
scale far below the GUT scale. This is because a see-saw mechanism
on the brane, with the right-handed neutrino masses being set by
this low fundamental scale, will typically not suppress the
associated contribution to the left-handed neutrino masses enough.
In this paper, we will not consider models with such
singlet neutrino states on the brane.

\medskip

In the next Section we present the general structure of the
five-dimensional brane-world model along with the
associated four-dimensional effective action. The Section concludes
with a discussion of the various symmetries that can be imposed on this
brane-world theory for model-building purposes. The spinor properties
used in this Section are
summarized in Appendix~\ref{app:spinor}. Section 3 contains a general
analysis of the structure of masses and mixing and a discussion
of the various resulting model building options. This discussion is
mainly based on a perturbative diagonalization of the mass matrix
assuming small brane-bulk mixing. The Section is complemented by
Appendix~\ref{app:diag} which contains some of the more technical
aspects related to the exact diagonalization of the mass matrix.
In Section 4 the constraints from energy loss in supernovae are
discussed. Section 5 describes a specific model with a bulk
Majorana mass which results in a small mixing between electroweak
eigenstates and bulk states. It is shown that, in the context of this
model, the small-angle MSW solution is a viable solution for the
solar neutrino problem. At the same time, the atmospheric oscillation
phenomena as well as the LSND signal can be understood.

\section{General framework}
\label{sec:general}

\noindent

In this Section, we will set up our formalism and present the general
structure of five-dimensional neutrino brane-world
models. Subsequently, we perform a Kaluza-Klein reduction to obtain
the associated four-dimensional effective theory which will be the
starting point for our analysis of neutrino masses and mixing.

\subsection{Five-dimensional brane-world models of neutrino physics}

\noindent

We consider brane world theories with a five-dimensional bulk
and coordinates denoted by $x^\a$ where $\a ,\b ,\cdots = 0,1,2,3,4$.
Spinors in this five-dimensional bulk theory provide
candidates for right-handed neutrinos and are, therefore, of
particular importance in our context. As mentioned above, within the
framework of string- or M-theory compactifications the presence of
such bulk spinors
is practically unavoidable, since they arise as superpartners of
moduli fields. We assume the existence of an arbitrary number, $N$, of
bulk Dirac spinors~\footnote{Our conventions for five-dimensional
spinors are summarized in Appendix~\ref{app:spinor}.} $\Psi_I$ where
$I,J,\cdots = 1,\cdots ,N$. Given these fields, one might want to
write the most general five-dimensional action bilinear in the spinors
and respecting five-dimensional Lorentz-invariance. However, there is
an important generalization to this that arises from including bulk
gauge fields $A_\a$ under which the fermions are charged. Within string- or
M-theory such bulk vector fields arise quite generically as part
of five-dimensional vector super-multiplets. Also gauging of the bulk
fermions with respect to these vector fields is realized in large
classes of models such as the five-dimensional brane-world models
originating from heterotic M-theory~\cite{Lukas:98b,Lukas:98c}.
Later on, we will compactify the five-dimensional theory on a certain
vacuum state to four dimensions. All we really have to require is that
this vacuum state respects four-dimensional Lorentz invariance. As a
consequence, in such a vacuum state, non-vanishing vacuum expectation
values (VEVs) for the components $A_4$ of the bulk vector fields in
the direction to be compactified are allowed. Hence, we should consider two
types of bulk mass terms. First, there are mass terms that respect
five-dimensional Lorentz invariance. We can think of such mass terms as being
generated by VEVs of bulk scalars. Secondly, there are mass terms
generated by VEVs of the components $A_4$ of bulk vector fields. They
arise once five-dimensional Lorentz invariance is spontaneously
broken to four-dimensional Lorentz invariance by the vacuum state.
We split the five-dimensional action as
\begin{equation}
 \label{S5}
 S_5=S_{\rm bulk}+S_{\rm brane}
\end{equation}
into a bulk and a brane part. Taking into account the two types of
mass terms as discussed, the relevant part of the bulk action is given
by
\bea 
S_{\rm bulk}
&=& \int d^5x\left[\bar{\Psi}_I\g^\a i\partial_\a\Psi_I
+\MD^S_{IJ}\bar{\Psi}_I\Psi_J-\frac{1}{2}(\MM^S_{IJ}
\bar{\Psi}^c_I\Psi_J+\text{h.c.})\right.\nonumber\\ &&
\left.+i\MD^V_{IJ}\bar{\Psi}_I\g_5\Psi_J
-\frac{1}{2}(\MM^V_{IJ}\bar{\Psi}^c_I\g_5\Psi_J+\text{h.c.})  \right]\; .
\label{Sbulk}
\eea
Here, $\MD_S$ and $\MM_S$ are the Lorentz-invariant
Dirac and Majorana mass matrices, respectively. The subscripts ``$S$''
refer to their possible origin as being generated by VEVs of bulk
scalar fields.  These mass matrices have two vector-like counterparts,
$\MD_V$ and $\MM_V$, generated by VEVs of bulk vector fields in the way
indicated above~\footnote{In the following, we will use
``$S$'' and ``$V$'' as upper or lower indices depending on convenience.}.
To see in some more detail how this happens, we note
that in five dimensions the spinors $\Psi$ and their charge conjugate
$\Psi^c$ transform in the same way under the five-dimensional Lorentz
group. Therefore, in addition to the ``ordinary'' kinetic terms of the
form $\bar{\Psi}\g^\a i\partial_\a\Psi$, we can also write Majorana
kinetic terms of the form $\bar{\Psi}^c\g^\a
i\partial_\a\Psi+\text{h.c.}$. In the most general case one should
gauge a linear combination of these terms. One can see that with an
appropriate redefinition in the $(\Psi ,\Psi^c)$ space the Majorana
kinetic term can always be removed. This is why, in the above action,
we have only written the standard kinetic term. However, in general,
one cannot simultaneously convert the couplings to
the vector field to the same standard form. In other words, once the
kinetic term is in the canonical form, the gauge transformations can
still mix the $\Psi$ and $\Psi^c$ fields~\footnote{As an example, one
can consider the case of one bulk family and a U(1) gauge group
acting on the fields by rotating the vector $(\Psi,\Psi^c)$.}.
Therefore, after taking into account the vector field VEVs, we indeed
generate the Dirac and the Majorana vector mass terms given
in~\eqref{Sbulk}. From the structure of the various terms
in~\eqref{Sbulk} we deduce the following symmetry properties of the
associated mass matrices.
\begin{equation}
 \label{mmsymm}
 \MD_S={\MD_S}^\dagger\; ,\qquad \MM_S={\MM_S}^T\; ,\qquad
 \MD_V={\MD_V}^\dagger\; ,\qquad \MM_V = {\MM_V}^T
\end{equation}

Are there other sensible generalizations of the bulk action that we
have not yet taken into account? In models where the fifth dimension
is divided by an orbifolding symmetry one should impose invariance of
the action only under those infinitesimal coordinate transformations
that are compatible with the orbifolding. Since this may be a smaller
set of transformations it leads to generalizations of the action given
above. Typically, in these cases, one can allow explicit
orbifold-dependent functions such as step-functions in the bulk
action. Another way of generating mass terms which effectively depend
on the additional dimension is to consider a vacuum state with a
non-flat metric. Explicit coordinate dependence of the mass terms
generated in these ways might have interesting consequences for the
structure of neutrino masses~\cite{neubert}. For example, generically
one expects that the Kaluza-Klein part of the mass matrix does not
have a block-diagonal structure any more. In this paper, we will not consider
this interesting possibility further but we remark that the general
formalism for the diagonalization of the mass matrix in
Appendix~\ref{app:diag} is equipped to handle even the more general
case of a non block-diagonal mass matrix.

Restricting to the case of coordinate-independent mass parameters,
the bulk Lagran\-gian~\eqref{Sbulk} constitutes the most general
expression compatible with four-dimensional Lorentz invariance and,
therefore, contains all possible mass terms. However, we can ask
whether we can generate these mass terms by mechanisms different
from the ones already mentioned. To do this, we note that the
spinors $\Psi_I$ transform in the fundamental of $Sp(4)\simeq SO(5)$.
A spinor bilinear, therefore, transforms under $Sp(4)$ as
\begin{equation}
 {\bf 4}\times{\bf 4}={\bf 1}+{\bf 5}+{\bf 10}\; .
\end{equation}
We have already discussed the coupling to the singlet ${\bf 1}$ and
the vector ${\bf 5}$ leading to the scalar and vector masses,
respectively. In addition, there can be a coupling to a second-rank
antisymmetric tensor field transforming in the representation
${\bf 10}$. However, only the components of this antisymmetric tensor
field pointing into the internal space can take VEVs if we are to
preserve four-dimensional Lorentz invariance. Clearly, for the case of
one additional dimension these components must vanish by antisymmetry
of the tensor field. Therefore, mass terms are generated by VEVs of
scalar and vector fields only.

\medskip

Next, we should set up the brane action. We choose $x^\m$, where $\m
,\n ,\cdots = 0,1,2,3$ as the coordinates longitudinal to the brane
and $y\equiv x^4$ as the transverse coordinate. Furthermore, we locate
the brane at $y=0$. The essential fields on the brane that we need to
consider for our purpose are the left-handed lepton doublets $L_i$,
where $i=e,\m ,\t$ and the relevant Higgs field $H$. As mentioned in
the introduction, we will not consider singlet neutrino fields on the
brane. We write the lepton doublets in a basis in family space where
the charged lepton mass matrix is diagonal and positive. The most
general brane-bulk Yukawa couplings invariant under the standard model
group can then be written as
\begin{equation}
 \label{Sbrane}
 S_{\rm brane} = \int_{\{ y=0\}}d^4x \left[-\frac{h_{Ii}}{\sqrt{M_5}}
                 \bar{\Psi}_IL_iH-\frac{h^c_{Ii}}{\sqrt{M_5}}
                 \bar{\Psi}^c_IL_iH+\text{h.c.}\right]\; .
\end{equation}
Since the bulk spinors with the normalization fixed by the bulk
action~\eqref{Sbulk} have mass dimension two, these couplings are
suppressed by the square root of a mass scale $M_5$ defined by
\begin{equation}
 \label{M5def}
 2\p RM_5 = \left(\frac{M_{\rm Pl}}{M_*} \right)^2
\end{equation}
where $M_*$ is the string scale. Furthermore, $R$ is the radius of the
circle on which we compactify the fifth dimension transverse to the
brane in preparation for our reduction to four dimensions. Also, we
restrict the range of the corresponding coordinate as $y\in [-\p R,\p
R]$ with the endpoints identified.

As mentioned in the introduction, in the context of string- or
M-theory, brane-bulk couplings may arise from higher-dimensional
gravity-induced operators after inserting appropriate VEVs. From a
theoretical viewpoint, their magnitude is rather uncertain and is, in
our case, parameterized by the dimensionless couplings $h_{Ii}$ and
$h^c_{Ii}$. In the context of string theory, we have, of course in
mind that our five-dimensional model is obtained by compactification
of the ten (or eleven) dimensional theory on a five- (or six-)
dimensional internal space with radii of compactification smaller than
$R$. In the simple case in which all those internal radii are equal
and given by $\rho$, the quantities $R$, $\r$ and the string scale
$M_*$ are related by
\begin{equation}
\label{stringscale}
 16\p (2\p RM_*)(2\p\r M_*)^5=\left(\frac{M_{\rm Pl}}{M_*}\right)^2\; .
\end{equation}
From this relation, in models with a low string scale $M_*$, we will
need $\rho^{-1}$ to be smaller than the string scale in order to get
the correct value for the low-energy Planck scale. In such cases, the
five-dimensional action specified by eq.~\eqref{S5} should be
considered as an effective theory below $\r^{-1}$ (rather than below
$M_*$). This also means~\footnote{We would like to thank R. Barbieri
for reminding us about this.} that we should restrict bulk masses to
be smaller than $\r^{-1}$.

If the bulk fermions propagate in all ten dimensions, as is suggested
by their possible interpretation as modulini, then, from \eq{M5def},
the dimensionless couplings $h_{Ii}$ and $h^c_{Ii}$ are simply given
by their ten-dimensional counterparts. On the other hand, if the bulk
fermions propagate only in $\d$ of the ten space-time dimensions
(where $5\leq\d <10$) the couplings $h$ are given by $h=h_\d (2\p\r
M_*)^{(10-\d )/2}$.

\subsection{The four-dimensional effective action}

\noindent

We would now like to reduce the above five-dimensional theory and
calculate the associated effective four-dimensional action. We require
the bulk fields to be periodic on the circle~\footnote{Other possible
choices are obtained by imposing generalized periodicity conditions of
Scherk-Schwarz type~\cite{Dienes:98a}. Those will lead to
complications in the structure of four-dimensional masses for the bulk
fields. Particularly, there can be a shift in the Kaluza-Klein
masses. However, we will not consider these exotic possibilities
here.}, that is
\begin{equation}
 \Psi_I(-\p R)=\Psi_I(\p R)\; .
\end{equation}
Correspondingly, the bulk fermions can be expanded into Kaluza-Klein
modes as
\begin{equation}
 \Psi_I(x,y) = \frac{1}{\sqrt{2\p R}}\sum_{n\in\Z}\Psi_{nI}(x)
               \exp\left(\frac{iny}{R}\right)\; .
\end{equation}
In addition, it is useful to express each of the modes $\Psi_{nI}$ in
terms of two four-dimensional (left-handed) Weyl spinors $\x_{-nI}$,
$\eta_{nI}$ as
\begin{equation}
 \Psi_{nI} =
 \begin{pmatrix}\bar{\x}_{-nI}\\\eta_{nI}\end{pmatrix}\; .
\end{equation}
Here the conjugate $\bar{\x}$ of a Weyl spinor $\x$ is defined as
$\bar{\x}=\e\x^*$ where $\e$ is the two-dimensional epsilon symbol.
For further details of our conventions we refer to
Appendix~\ref{app:spinor}. The four-dimensional effective action associated
to the brane-world theory~\eqref{S5}, \eqref{Sbulk}, \eqref{Sbrane}
is then given by
\begin{equation}
 \label{L}
 \cL = \cL_{\rm kin}+\cL_{\rm mass}
\end{equation}
where the kinetic terms of the bulk fields read
\begin{equation}
 \cL_{\rm kin} = \sum_{n\in\Z}\left[\eta^\dagger_{nI}\s^\m i\partial_\m
                 \eta_{nI}+\x^\dagger_{nI}\s^\m i\partial_\m
                 \x_{nI}\right]\; .
\end{equation}
The all-important mass part of the Lagrangian can be conveniently
written in the form
\begin{equation}
 \label{Lmass}
 \cL_{\rm mass} = -\sum_{n\in\Z}\left[\frac{1}{2}
                  \begin{pmatrix}\eta_n^T&\x_n^T\end{pmatrix}
                  \cM_n\begin{pmatrix}\eta_{-n}\\\x_{-n}
                  \end{pmatrix}+\begin{pmatrix}\eta_n^T
                  &\x_n^T\end{pmatrix}\cN\n\right]+\text{h.c.}
\end{equation}
where, for ease of notation, we have adopted a vector notation in
family space. That is, for example, $\eta_n=(\eta_{n1},\cdots
,\eta_{nN})^T$ and $\n = (\n_e,\n_\m ,\n_\t )^T$. Given our
conventions for Weyl mass terms in Appendix~\ref{app:spinor}, here, the
transpose $T$ refers to family space only. Written in
block-form, the mass matrix $\cM_n$ for the Kaluza-Klein modes
at level $n$ takes the form
\begin{equation}
 \label{Mn}
 \cM_n = \begin{pmatrix}\MM_1&\MD^T-\frac{in}{R}\\
         \MD +\frac{in}{R}&\MM_2\end{pmatrix}\; .
\end{equation}
Here $\MM_1$ and $\MM_2$ are Majorana mass-matrices related to
the scalar and vector Majorana mass-matrices appearing in the
five-dimensional bulk action~\eqref{Sbulk} by
\begin{equation}
 \MM_1 = -\MM_S+\MM_V\; ,\qquad \MM_2=(\MM_S+\MM_V)^*\; .
\end{equation}
The Dirac mass-matrix $\MD$ encodes both scalar and vector Dirac masses
and is explicitly defined by
\begin{equation}
 \MD = \MD_S-i\MD_V\; .
\end{equation}
{}From eq.~\eqref{mmsymm}, it is clear that $M_1$ and $M_2$ are
symmetric matrices while, generally, there is no further constraint on
$\m$. 

Note that the bulk mass matrices satisfy the relations
\begin{equation}
 \cM_{-n}=\cM_n^T\; .
\end{equation}
They ensure that the Majorana mass matrix 
\bea
\renewcommand{\arraycolsep}{0.3cm}
&&\qquad\qquad\;\,\begin{array}{llllll}\n^T&\eta_0^T\;\x_0^T&\cdots&
\eta_{-n}^T\;\x_{-n}^T&
\eta_{n}^T\;\x_{n}^T&\cdots\end{array} \nonumber\\[2mm]
\cM &=&
\begin{array}{c}\n\\\begin{array}{c}\eta_0\\\x_0\end{array}\\
                \vdots\\\begin{array}{c}\eta_{-n}\\\x_{-n}
                \end{array}\\\begin{array}{c}\eta_n\\\x_n\end{array}\\
                \vdots\end{array}
\left(\begin{array}{llllll}
           \phantomheigth{\n}{0}&\phantombox{$\eta_0^T\;\x_0^T$}
           {$\cN^T$}&\cdots&
           \phantombox{$\eta_{-n}^T\;\x_{-n}^T$}{$\cN^T$}&
           \phantombox{$\eta_{n}^T\;\x_{n}^T$}{$\cN^T$}&\cdots\\
           \phantomheigth{\begin{array}{l}\eta_0\\\x_0\end{array}}{\cN}
           &\phantombox{$\eta_0^T\;\x_0^T$}{$\cM_0$}&\cdots&
           \phantombox{$\eta_{-n}^T\;\x_{-n}^T$}{$0$}&
           \phantombox{$\eta_{n}^T\;\x_{n}^T$}{$0$}&\hdots\\
           \vdots&\phantombox{$\eta_0^T\;\x_0^T$}{$\vdots$}&
           \ddots&\phantombox{$\eta_{-n}^T\;\x_{-n}^T$}{$0$}&
           \phantombox{$\eta_{n}^T\;\x_{n}^T$}{$0$}&\hdots\\
           \phantomheigth{\begin{array}{c}\eta_{-n}\\\x_{-n}
           \end{array}}{\cN}&\phantombox{$\eta_0^T\;\x_0^T$}{$0$}&0&
           \phantombox{$\eta_{-n}^T\;\x_{-n}^T$}{$0$}&
           \phantombox{$\eta_{n}^T\;\x_{n}^T$}{$\cM_n^T$}&\\
           \phantomheigth{\begin{array}{c}\eta_n\\\x_n\end{array}}{\cN}
           &\phantombox{$\eta_0^T\;\x_0^T$}{$0$}&0&
           \phantombox{$\eta_{-n}^T\;\x_{-n}^T$}{$\cM_n$}&
           \phantombox{$\eta_{n}^T\;\x_{n}^T$}{$0$}&\\
           \vdots&\phantombox{$\eta_0^T\;\x_0^T$}{$\vdots$}&\vdots&&&\ddots
\end{array}\right) \label{mmatrix}
\renewcommand{\arraystretch}{1}
\eea
associated with the mass Lagrangian~\eqref{L} is indeed symmetric as it
should be.

The brane-bulk coupling matrix $\cN$ has the form
\begin{equation}
 \cN = \begin{pmatrix}\mbbc\\\mbb\end{pmatrix}\; .
\end{equation}
The mass matrices $\mbb$ and $\mbbc$ are related to the 
brane-bulk couplings in the brane action~\eqref{Sbrane},
respectively and are explicitly given by
\begin{equation}
 \label{branebulk}
 \mbb_{Ii} = \frac{vh_{Ii}}{\sqrt{2\p RM_5}}=h_{Ii}v\frac{M_{*}}
 {M_{\rm Pl}}\; ,\qquad
 \mbbc_{Ii} = \frac{v h^c_{Ii}}{\sqrt{2\p R\MM_5}}=h_{Ii}^cv\frac{M_{*}}
 {M_{\rm Pl}}\; .
\end{equation}
Here $v$ is the Higgs vacuum expectation value.

In what follows we will have occasion to choose the brane-bulk
couplings, $h_{Ii}$, to be of a given size for phenomenological
reasons. However, there are lower bounds on $M_*/h_{Ii}$
at tree level or on $M_*/h_{Ii}^2$ at one-loop-level by requiring that
there be no significant departure from Standard Model
processes~\cite{ioannisian}. It turns out that these constraints
can be satisfied within the explicit model to be discussed later.
Moreover, it has been suggested that values of $h_{Ii}$ larger than one
may be inconsistent with staying in the perturbative
domain\footnote{We thank J. Valle, A. Ioannisian and A. Pilaftsis for
comments on this point.}. However, from \eq{stringscale}, we
see that the string scale is very sensitive to the nature of
compactification from higher dimensions so it is always possible to
arrange for $h_{Ii}$ to remain perturbative through a
choice of $M_*$.

\subsection{Symmetries of the brane-world action}

\noindent

For the purpose of model building it is important to consider the
various symmetries that can be imposed on the general brane-world
action~\eqref{S5} in order to forbid some of the mass terms. A general
classification of these symmetries can be obtained by analyzing 
the invariance properties of the bulk kinetic terms. Instead of
going through such a systematic procedure, here we merely list 
some of the most important symmetries. The most obvious
candidate is the part of the five-dimensional Lorentz group connected
to the group identity which we call ${\cal L}_0$. Infinitesimal
transformations of this type act on the fields in our model as
\begin{equation}
 \d\Psi_I=\frac{i}{2}\o^{\a\b}\Sigma_{\a\b}\Psi_I\; ,\qquad
 \d L_i = \frac{i}{2}\o^{\m\n}\Sigma_{\m\n}L_i
\end{equation}
where $\o^{\a\b}$ are infinitesimal parameters and $\Sigma_{\a\b}$ are the
generators defined in Appendix~\ref{app:spinor}. The original
five-dimensional action should, of course, be invariant under this
symmetry. However, as discussed above, this is not necessarily the case for
the vacuum state of the five-dimensional theory. Imposing this
symmetry, therefore, means requiring a vacuum state which is
invariant under infinitesimal five-dimensional Lorentz
transformations. Correspondingly, this symmetry forbids the two
vector-like bulk mass terms in the action~\eqref{Sbulk} that we have
attributed to spontaneous Lorentz-symmetry breaking by the vacuum
state. On the other hand, it allows all other terms.

Furthermore, we can consider the generators of those parts of the
five-dimensional Lorentz group that are not connected to the identity. An
important example is the parity transformation
\begin{equation}
 \label{P5}
 P_5\; :\quad y\rightarrow -y
\end{equation}
in the direction transverse to the brane. On the fields it acts as
\begin{equation}
 \label{P5a}
 \Psi_I(y)\rightarrow\g_5\Psi_I (-y)\; ,\qquad L_i\rightarrow L_i\; .
\end{equation}
In principle, the transformation law for $\Psi_I$ has a sign
ambiguity. However, $\Psi_I$ and its charge conjugate $\Psi_I^c$
transform with opposite signs. We can therefore remove this sign
ambiguity by exchanging $\Psi_I$ and $\Psi_I^c$ appropriately and
calling $\Psi_I$ those fields that transform with a
positive sign under $P_5$ as in eq.~\eqref{P5a}. It is also
useful to have the action of this symmetry on the Weyl Kaluza-Klein
states appearing in the four-dimensional effective action available.
It is given by
\begin{equation}
 \label{P5action}
 \x_{nI}\rightarrow \x_{-nI}\; ,\qquad \eta_{nI}\rightarrow
  -\eta_{-nI}\; ,\qquad
 \n_i\rightarrow \n_i\; .
\end{equation}
This $\ZZ$ symmetry forbids bulk Dirac mass terms and the
brane-bulk couplings $\mbb^c$. It allows all other mass terms.

For $N$ bulk spinors the bulk kinetic terms have a global $U(N)$
symmetry. Particularly interesting for model building are the $U(1)$
subgroups acting on the spinors as
\begin{equation}
 \Psi_I\rightarrow e^{i\a Q_I}\Psi_I\; ,\qquad
 L_i\rightarrow e^{i\a q_i}L_i
\end{equation}
where $\a$ is a continuous parameter and $Q_I$ and $q_i$ are charges.
The group action on the Kaluza-Klein modes reads
\begin{equation}
 \x_{nI}\rightarrow e^{-i\a Q_I}\x_{nI}\; ,\qquad
 \eta_{nI}\rightarrow e^{i\a Q_I}\eta_{nI}\; ,\qquad
 \n_i\rightarrow e^{i\a q_i}\n_i\; .
\end{equation}
A specific example is lepton number defined by $Q_I=q_i=1$. This
symmetry allows the Dirac mass terms and forbids the Majorana mass
terms. Clearly there are many more interesting choices that 
can be made in order to constrain the mass terms in a sensible way.
We will encounter another example later on. In the context of string-
or M-theory, if one thinks about these symmetries as being global,
it appears more likely to have invariance under a discrete
subgroup of $U(1)$ rather than under the full continuous symmetry.
Hence, when we consider $U(1)$ symmetries in the following, we will
always have in mind that one might have to replace those symmetries
by appropriate discrete subgroups.

In Table~\ref{table:symmetries} we have listed the various mass terms
and their transformation properties under the above symmetries in
order to give an overview of the constraints available for model-building. 
\begin{table}
 \begin{center}
 \begin{tabular}{|l||l|l|l|}\hline
  &$\cL_0$&$P_5$&$U(1)$\\\hline\hline
  scalar Dirac $\MD_{IJ}^S$&inv.&not inv.&$Q_I=Q_J$\\\hline
  scalar Majorana $\MM_{IJ}^S$&inv.&inv.&$Q_I=-Q_J$\\\hline
  vector Dirac $\MD_{IJ}^V$&not inv.&not inv.&$Q_I=Q_J$\\\hline
  vector Majorana $\MM_{IJ}^V$&not inv.&inv.&$Q_I=-Q_J$\\\hline
  brane-bulk $\mbb_{Ii}$&inv.&inv.&$Q_I=q_i$\\\hline
  brane-bulk $\mbbc_{Ii}$&inv.&not inv.&$Q_I=-q_i$\\\hline
 \end{tabular}
 \end{center}
 \caption{\em Invariance properties of the mass terms with respect to
           the symmetries introduced in the text. For the $U(1)$ symmetries
           the conditions for invariance are stated.}
 \label{table:symmetries}
\end{table}

A final remark concerns models that are obtained by compactification
on the orbifold $S^1/\ZZ$ where the action of the $\ZZ$ symmetry is
specified by $P_5$. While, so far, we have just considered the
reduction on a circle, the orbifold case does, in fact, not need to be treated
separately for our purposes. This is for the following reason. We can
consider our general four-dimensional model specified by the
Lagrangian~\eqref{L} and impose $P_5$ as a symmetry. Using the
$P_5$ transformation properties~\eqref{P5action} of the
four-dimensional fields, we can then define $\ZZ$ even and odd modes.
Our Lagrangian~\eqref{L}, being bilinear in the fields, then
decomposes into two parts, one which involves only even fields the
other one involving only odd fields. The ordinary neutrinos $\n_i$ are
even fields and, therefore, only appear in the even part of the
Lagrangian. As a consequence, the odd modes decouple from the physics
that we are considering in this paper. Orbifolding such a $\ZZ$
invariant model merely means that the already decoupled odd states
are truncated. We therefore see that for each orbifold model there
is a corresponding $P_5$ invariant model with equivalent physics as
far as neutrino masses and mixing are concerned.
 
\section{General structure of masses and mixing}

\noindent

It is clearly desirable to obtain general results about the masses and
mixing in our brane-world model that are independent of
specific model-building considerations. We would therefore like to
present a number of general conclusions derived from the effective
four-dimensional mass Lagrangian~\eqref{Lmass} before discussing explicit
models.

\subsection{Masses of bulk modes}

\noindent

As a first step, we will analyze the bulk part of the mass
Lagrangian separately. This bulk part is given by the first term in
eq.~\eqref{Lmass} and reads
\begin{equation}
 \label{Lbulk}
 L_{\rm bulk} =  -\frac{1}{2}\sum_{n\in\Z}
                  \begin{pmatrix}\eta_n^T&\x_n^T\end{pmatrix}
                  \cM_n\begin{pmatrix}\eta_{-n}\\\x_{-n}
                  \end{pmatrix}+\text{h.c.}
\end{equation}
where the mass matrices $\cM_n$ have been defined in eq.~\eqref{Mn}.
Clearly, the mass diagonalization of this bulk action will have to
be corrected when the brane-bulk couplings are taken into account.
However, usually we will be interested in the case where these
couplings and, hence, the corresponding corrections are small. In this case,
studying~\eqref{Lbulk}, provides us with useful information about the
approximate spectrum of bulk states.

\medskip

Let us first note that the mass matrices $\cM_n$ can be diagonalized
by unitary matrices $U_n$, where $n\in\Z$, such that
\begin{equation}
\label{Mnhat}
\hat{\cM}_n = U_n^T\cM_n U_{-n}
\end{equation}
are diagonal matrices with entries
\begin{equation}
 \label{Mneigen}
 \hat{\cM}_n=\text{diag}(M_{n1+},\cdots ,M_{nN+},M_{n1-},\cdots ,M_{nN-})\; .
\end{equation}
The original states $(\eta_n ,\x_n)$ are related to the mass
eigenstates $(\hat{\eta}_n,\hat{\x}_n)$ by
\begin{equation}
 \begin{pmatrix}\eta_n\\\x_n\end{pmatrix}
  = U_n \begin{pmatrix}\hat{\eta}_n\\\hat{\x}_n\end{pmatrix}\; .
\end{equation}
The eigenvalues and the matrices $U_{-n}$ (up
to unitary rotations within each eigenspace) can be computed by
diagonalizing the hermitian matrices $\cM_n^\dagger\cM_n$.

For each $n>0$ we have $4N$ Weyl mass eigenstates
$(\hat{\eta}_n,\hat{\eta}_{-n},\hat{\x}_n,\hat{\x}_{-n})$. It is
immediately clear from the structure of~\eqref{Lbulk} that the
components with opposite mode number $n$ arrange themselves into
Dirac states. For each integer $n>0$, we, therefore, have $2N$ Dirac
states with masses denoted by $(M_{nI+},M_{nI-})_{I=1,\cdots ,N}$.
This result was, perhaps, not quite so obvious from the original
five-dimensional bulk action~\eqref{Sbulk}. We remark that the corrections
induced by the brane-bulk couplings (to be included below) will generally not
respect this Dirac structure. However, for small couplings we will still
retain approximate Dirac states. The modes for $n=0$, on the other hand,
do not generally group themselves into Dirac states and represent
$2N$ Weyl states with Majorana-Weyl masses denoted by
$(M_{0I+},M_{0I-})_{I=1,\cdots ,N}$.

\medskip

It is difficult to extract more information about the spectrum of bulk
states without further assumptions. To see the range of possibilities,
we will, therefore, consider two specialized cases. First, we
consider a situation where the vector mass matrices are zero, that is
$\MD_V=\MM_V=0$, and the scalar mass matrices $\MD_S$, $\MM_S$ are real.
Then we have
\begin{equation}
 \label{Mc1}
 \cM_n = \begin{pmatrix}-\MM_S&\MD_S-\frac{in}{R}\\
        \MD_S+\frac{in}{R}&\MM_S\end{pmatrix}\; .
\end{equation}
Hence, the matrix $\cM_0$ for the lowest mode $n=0$ is symmetric and
can be diagonalized by an orthogonal matrix $V$ leading to
$\hat{\cM}_0=\text{diag}(M_{0I+},M_{0I-})_{I=1,\cdots ,N}=V^T\cM_0V$. 
Then one can easily see that the matrices $U_n$ defined by
\begin{equation}
 \label{Un1}
 U_n=\left\{\begin{array}{lll}V&\text{for}&n=0\\
     PV&\text{for}&n>0\\
     P^TV&\text{for}&n<0\end{array}\right.
\end{equation}
with
\begin{equation}
 P = \frac{1}{\sqrt{2}}\left(\begin{array}{rr}
     \um_N&\um_N\\-\um_N&\um_N\end{array}\right)
\end{equation}
diagonalize the mass matrices $\cM_n$ for arbitrary mode number $n$.
{}From eqs.~\eqref{Mc1} we then find for the eigenvalues (neglecting
phase factors)
\begin{equation}
 \label{spec1}
 M_{nI\pm} = \sqrt{M_{0I\pm}^2+\frac{n^2}{R^2}}\; .
\end{equation}
We recall that, in this equation, $M_{0I\pm}$ are the masses of the
$n=0$ modes which depend on the scalar Dirac and Majorana mass
matrices. We can therefore express the complete mass spectrum as well as
the diagonalizing unitary matrices in terms of the corresponding
quantities for the $n=0$ mode. From the general discussion above it is
clear that the $2N$ masses given by eq.~\eqref{spec1} correspond
to $2N$ Weyl states for $n=0$ and to $2N$ Dirac states for $n>0$.
The spectrum~\eqref{spec1} has the
form one usually expects for a Kaluza-Klein tower of particles.
In particular, the masses $M_{0I\pm}$ and, hence, the scalar Dirac and
Majorana masses, provide a lower cutoff for the spectrum. Therefore,
in order to have light bulk masses, the scalar mass terms must be
small compared to the string scale. As discussed earlier, this is the
natural expectation if the bulk fermions are interpreted as
superpartners of moduli fields. As we will see below, the situation is
quite different for the vector-like masses as they do not represent
lower bounds on the bulk mass spectrum.

\medskip

In our second example, we would like to include all types of mass
terms. In particular, we are interested in the effect of the
vector-like masses which we have previously set to zero.
As a simplification, we restrict the analysis to the case of one
family, that is $N=1$ and assume that all mass parameters are real.
We are, therefore, dealing with $2\times 2$
mass matrices
\begin{equation}
 \cM_n = \begin{pmatrix}-\MM_S+\MM_V&\MD_S-i\MD_V-\frac{in}{R}\\
         \MD_S-i\MD_V+\frac{in}{R}&\MM_S+\MM_V\end{pmatrix}\; .
\end{equation}
In this case, the two eigenvalues of $\cM^\dagger_n\cM_n$ are given by
\begin{equation}
 \label{spec2}
 M_{n\pm}^2 = \MM_S^2+\MD_S^2+\MM_V^2+\MD_V^2+\frac{n^2}{R^2}\pm 2
             \sqrt{\MD_V^2\left(\frac{n^2}{R^2}+\MM_S^2\right)
             +\MM_V^2\left(\frac{n^2}{R^2}+\MM_S^2+\MD_S^2\right)}\; .
\end{equation}
The first five terms on the left-hand side of this equation are as
expected and are of the form~\eqref{spec1} that we have encountered in
the absence of vector-like masses. The square root term (which
vanishes for vanishing vector-like masses) does not fit into this
structure. Specifically the eigenvalues corresponding to the negative
sign in front of the square root are not bounded from
below by the vector-like masses. To see this in more detail let us
consider the four cases in which only one type of mass term is present.
For a non-vanishing scalar Dirac mass but all other mass terms vanishing
we have
\begin{equation}
 M_{n\pm}^2 = \MD_S^2+\frac{n^2}{R^2}\; .
\end{equation}
The spectrum is of the form \eq{spec1} and is indeed bounded from
below by $\MD_S$. Similarly, this is the case for a scalar Majorana
mass and all other mass terms vanishing. In this case we have
\begin{equation}
 M_{n\pm}^2 = \MM_S^2+\frac{n^2}{R^2}\; .
\end{equation}
The situation is quite different, however, for the vector-like mass
terms. Let us consider a situation with a vector Dirac mass only.
Then the spectrum is given by
\begin{equation}
 M_{n\pm}^2 = \left(|\MD_V|\pm\frac{|n|}{R}\right)^2\; . 
\end{equation}
Analogously, for a vector Majorana mass as the only mass term we have
\begin{equation}
 \label{Mspec}
 M_{n\pm}^2 = \left(|\MM_V|\pm\frac{|n|}{R}\right)^2\; . 
\end{equation}
Hence, in these last two cases, the spectrum is not bounded from below
by the vector-like bulk mass term. While the mass of the $n=0$ mode is
still given by the vector bulk mass (as in the scalar case) the masses of some
of the higher modes are lower. In fact, the modes $n$ specified by the
integer part of $\pm\MD_SR$ ($\pm\MM_SR$) have the smallest mass given
by $\MD_S$ modulo $R^{-1}$ ($\MM_S$ modulo $R^{-1}$). As a consequence,
there will always be modes with masses of the order $1/R$ independent on
the size of the vector bulk masses. 

\subsection{Including brane-bulk couplings}

\label{bbcouplings}
\noindent

So far, we have analyzed the spectrum of the bulk modes for vanishing
brane-bulk coupling. However, our main task is to diagonalize the full
mass matrix~\eqref{mmatrix}. A general formalism for an exact
diagonalization is set up in Appendix~\ref{app:diag}.
However, in many cases it is sufficient to use a perturbative
diagonalization leading to a see-saw type mechanism. This allows
one to bring the matrix~\eqref{mmatrix} into block-diagonal form.
The approximation is also well-suited to understand the physical
implications of our models intuitively. To see how this
works, it is useful to introduce the quantities
\begin{equation}
 \e_n = \cM_n^{-1}\cN\; .
\end{equation}
For a see-saw diagonalization to be sensible these quantities should
be small compared to one. However, we should also take into account that we
have to sum over a large number of modes $n$. Therefore, the more
precise requirement is that
\begin{equation}
 \label{e2}
 \e^2=\sum_{n\in\Z}\e_n^\dagger\e_n
\end{equation}
should be small compared to one. Provided this is the case, we can 
perturbatively diagonalize the mass matrix neglecting terms of order
$\e^2$ and higher. To do this, we define the ``see-saw'' basis
$(\n ',\x_n',\eta_n')$ by
\begin{align}
\label{basis}
 \begin{pmatrix}\eta_n'\\\x_n'\end{pmatrix} &=
 \begin{pmatrix}\eta_n\\\x_n\end{pmatrix}+\e_n\n \mytag \\
 \n ' &= \n - \sum_{n\in\Z}\e_{-n}^\dagger
 \begin{pmatrix}\eta_n\\\x_n\end{pmatrix} \mytag \; .
\end{align}
Diagonalizing, the mass Lagrangian~\eqref{Lmass} then takes the
block-diagonal form
\begin{equation}
 \label{Lmassp}
 \cL_{\rm mass} = -\frac{1}{2}{\n '}^T\ml\n '-\frac{1}{2}\sum_{n\in\Z}
                   \begin{pmatrix}{\eta_n'}^T&
                  {\x_n'}^T\end{pmatrix}\cM_n\begin{pmatrix}
                  \eta_{-n}'\\\x_{-n}'\end{pmatrix}\; .
\end{equation}
Here, the mass matrix $\ml$ for the light neutrinos is given by
\begin{equation}
\label{seesaw}
 \ml = -\sum_{n\in\Z}\cN^T\cM_n^{-1}\cN\; .
\end{equation}
These results hold to order $\e$ and receive corrections at order
$\e^2$. In a second step we now need to diagonalize the various blocks
in \eqref{Lmassp}. To do this, we introduce the mass eigenstate basis
$(\hat{\n},\hat{\eta}_n,\hat{\x}_n)$ related to the see-saw basis by
unitary rotations
\begin{equation}
 \n ' = U\hat{\n}\; ,\qquad
 \begin{pmatrix}\eta_n'\\\x_n'\end{pmatrix}=
  U_n \begin{pmatrix}\hat{\eta}_n\\\hat{\x}_n\end{pmatrix}
\end{equation}
such that the mass matrices
\begin{equation}
 \hat{m} = U^T\ml U\; ,\qquad \hat{\cM}_n=U_n^T\cM_nU_{-n}
\end{equation}
are diagonal. The diagonalization of the bulk states is exactly as
discussed in the previous Subsection and the unitary matrices $U_n$
are  precisely as introduced in eq.~\eqref{Mnhat}. We can, therefore,
use the results for the bulk mass spectrum and the matrices $U_n$
obtained there. The diagonalization of the light mass matrix $\ml$ is
clearly more complicated and will, for particular cases, be
explicitly carried out later on. With these quantities available
we can now express the electroweak eigenstates $\n$ in terms of the
mass eigenstates $(\hat{\n},\hat{\eta}_n,\hat{\x}_n)$ as
\begin{equation}
 \label{nueigen}
 \n = U\hat{\n}+\sum_{n\in\Z}\e_{-n}^\dagger U_n\begin{pmatrix}
      \hat{\eta}_n\\\hat{\x}_n\end{pmatrix}
\end{equation}
where we have neglected terms of order $\e^2$. This equation can, of
course, also be derived by expanding the exact
formula~\eqref{eigenvectors} to first order in $\e$. The
result~\eqref{nueigen} is, in
fact, central to the subsequent discussion of the oscillation
phenomenology. Provided that the quantity $\e^2$ defined
in eq.~\eqref{e2} is much smaller than one, as we have assumed, this
equation asserts that the electroweak eigenstates are predominantly
composed of light mass eigenstates $\hat{\n}$ but have small
admixtures of the heavy mass eigenstates $(\hat{\eta}_n,\hat{\x}_n)$.
The size of these admixtures is determined by the quantities $\e_n$.
The mixing matrix $U$ then plays the role of the standard
(MNS) neutrino mixing matrix. This offers the interesting possibility that
some of the observed oscillation phenomena could be attributed to
oscillations into the light states (playing the role of the standard
neutrinos) while other phenomena could be due to oscillations into the
heavy states (which are predominantly bulk states). In fact,
given the various mass terms in our model, the parameters in these
two sectors can be largely independent.

\medskip

To verify this last statement, it is useful to evaluate some of the
above quantities more explicitly. We introduce the matrices
\bea
 X_n &=& \left[\MD^T-\frac{in}{R}-\MM_1\left(\MD +\frac{in}{R}\right)^{-1}
         \MM_2\right]^{-1} \\
 R_n &=& \left(\MD +\frac{in}{R}\right)^{-1}\MM_2X_n \\
 S_n &=& X_n\MM_1\left(\MD +\frac{in}{R}\right)^{-1}
\eea
and further note that $R_{-n}=R_n^T$ and $S_{-n}=S_n^T$. With these
definitions we find for the mixing parameters $\e_n$ of the bulk states
\begin{equation}
 \e_n = \begin{pmatrix}-R_n\mbbc+X_{-n}^T\mbb\\
        X_n\mbbc-S_n\mbb\end{pmatrix}\; .
\end{equation}
In addition, the mass matrix $\ml$ for the light states is given by
\begin{equation}
 \label{mp}
 \ml = \sum_{n\in\Z}\left[{\mbbc}^TR_n\mbbc+\mbb^TS_n\mbb-
       {\mbbc}^TX_n^T\mbb-\mbb^TX_n\mbbc\right]\; .
\end{equation}
Unfortunately, it is difficult to explicitly sum this series in the
most general case. However, partial results for special cases will be
given below.
We would now like to specialize these results to models that preserve
standard lepton number~\cite{LR,LRRR} with charges $Q_I=q_i=1$. From
Table~\ref{table:symmetries}, we see that in this case all
Majorana-type mass terms, that is $\MM_S$, $\MM_V$ and $\mbbc$ must
vanish. As a consequence, we have $R_n=S_n=0$ and $\ml = 0$.
Therefore, we have three massless Majorana states. Being a direct
consequence of imposing the lepton number symmetry these states are
exactly massless even beyond the see-saw approximation that we are
considering here. The mixing parameters take the simple form
\begin{equation}
 \e_n = \begin{pmatrix} \left(\MD +\frac{in}{R}\right)^{-1}\mbb\\0
        \end{pmatrix}\; .
\end{equation}
{}From eq.~\eqref{nueigen} we, therefore, see that the electroweak
eigenstates are generally non-trivial superpositions of the massless
states and the bulk states. The mixing matrix $U$ in
eq.~\eqref{nueigen} is, of course, arbitrary and corresponds to a
redefinition of the degenerate massless states. The size of the bulk
components $n$ in this superposition is controlled by the quantity
$(\MD +in/R)^{-1}\mbb$. For Dirac masses $\MD\gg \mbb$ these bulk
components are small and the electroweak eigenstates are predominantly
given by the massless states. This has to be contrasted with the
opposite case $\MD\ll \mbb$ or rather $\MD =0$ that has been mostly
considered in the literature so far. In this case, the massless states
are simply the zero modes of the Kaluza-Klein towers and they decouple
from the electroweak eigenstates. Moreover, the masses of the lightest
modes are then given by the brane-bulk mixing $m$. 
In contrast, in the presence of a bulk Dirac term, the masses of the
lightest bulk modes are determined by this Dirac mass. Keeping the Dirac
mass fixed, we can then vary the  brane-bulk coupling $m$ thereby
changing the importance of the heavy components in the electroweak
eigenstates and, at the same time, keeping three massless,
non-decoupled states. Hence, the lepton-number conserving mass
terms control the oscillation phenomenology associated with the
bulk states. Small lepton-number breaking Majorana-type terms,
on the other hand, generate small masses for the formerly massless
states and fix the mixing matrix $U$ in eq.~\eqref{nueigen}.
Therefore, they control the oscillation phenomena related to the
light states. This confirms the picture in which there are two
different sectors, both potentially relevant for oscillation
phenomenology, and which, by choosing the various parameters in
the model, can be controlled quite independently. As we will see,
to some extend this is also the case for the model with bulk Majorana
terms to be considered in Section~\ref{sec:5} although there the
Majorana term determines the mass of the lightest bulk states
and is also involved in the expression for the light neutrinos. 

\subsection{Model building options and scales}

\noindent

We have seen that our models typically accommodate two sectors of
mass eigenstates, both of which generically couple to the electroweak
eigenstates. In particular, there are three light eigenstates
which (in the see-saw approximation) are comparable to the mass
eigenstates in standard three-neutrino models. In addition, we have
Kaluza-Klein towers of heavy states. Oscillations into both
sectors may play a role in explaining the observed oscillation
phenomena. A broad classification of the various possibilities
can be given using the relevance of the Kaluza-Klein states for
oscillation phenomenology as a criterion. The relevance of the
Kaluza-Klein sector is basically controlled by the compactification
scale $1/R$ and the smallest mass $M_{\rm min}$ in the bulk
Kaluza-Klein spectrum~\footnote{In cases with vanishing
vector-like bulk masses this smallest mass coincides with the mass
of the $n=0$ mode. For non-vanishing vector-like masses, however,
this is not necessarily the case. See the discussion around
eq.~\eqref{spec2} for further details.}. These
should be compared to the mass differences, $\D
m$, needed to explain the various oscillation phenomena. Normally, then,
we take $\D m$ to be in the range relevant for either the MSW solution
of the solar neutrino problem, that is,
$\D m = 10^{-3}\text{--} 10^{-2}\text{eV}$ or in the range for
oscillations of atmospheric neutrinos, that is,
$\D m = 10^{-2}\text{--} 10^{-1}\text{eV}$. We can now distinguish
three cases.
\begin{itemize}
 \item $1/R\gg \D m$ and $M_{\rm min}\gtrsim\D m$~: In this case, none of
 the Kaluza-Klein states play a direct role for the observed
 oscillation phenomena. Therefore, these phenomena have to be
 explained entirely by oscillations between the standard neutrinos,
 that is, by the first term in eq.~\eqref{nueigen}. In such a
 situation, it may  seem that the brane-world nature of the model
 does not play any role  at all. However, as can be seen from eq.~\eqref{mp},
 these states do affect the mass matrix of the light neutrinos and,
 hence, the standard mixing matrix $U$. As we will discuss in more
 detail below,  we can have  models which mimic ordinary see-saw
 models with the  mass scale of the right-handed neutrinos set by
 a bulk mass term or models with ``geometrical'' see-saw
 mechanism~\cite{Dienes:98a}  where the right-handed mass scale is given
 by $1/R$. In summary, this case may, therefore, lead to interesting
 alternatives to conventional ``see-saw'' models of neutrino masses.

\item $1/R\gg\D m$ and $M_{\rm min}\lesssim\D m$~: In this case, only the bulk
 modes with the minimal mass $M_{\rm min}$ are directly accessible
 for oscillations. Such models are, therefore, equivalent to
 conventional models containing a (small) number of sterile neutrinos.
 We remark that an explanation of the atmospheric neutrino deficit
 due to oscillation into a sterile neutrino appears to be
 disfavored~\cite{Sobel2000}. Given this one must try to use this
 case to explain the solar neutrino deficit by sterile neutrinos
 and the atmospheric deficit (and, perhaps, the LSND result) by
 standard oscillations. However, large mixing angles between
 active and sterile neutrinos might be problematic from the
 viewpoint of supernova physics as well as standard model
 universality constraints~\cite{ioannisian}. On the
 other hand, the small mixing angle solution for oscillation into a
 sterile neutrino seems to be disfavoured by recent Super-Kamiokande
 data~\cite{Suzuki2000}. Hence, this case, although perhaps not
 excluded does not appear to be very attractive.

\item $1/R\lesssim\D m$ and $M_{\rm min}\lesssim\D m$~: In this case,
 a number of bulk modes can contribute to oscillation phenomena. The
 case that $\n_e$ or $\n_\m$ has a large bulk component in order to
 realize the large angle MSW solution is problematic for the same
 reasons as above. On the other hand if the mixing of bulk modes with
 the electroweak eigenstates is small then the small-angle MSW solution
 of the solar neutrino problem might be attributed to oscillations into
 bulk states. As we will discuss, the phenomenology is somewhat
 different from the second case and, in fact, compatible with the
 recent Super-Kamiokande results as well as with the other solar
 neutrino experiments. At the same time, the other oscillation
 phenomena should then be explained by oscillations between the
 standard neutrinos.

\end{itemize}

Some of these options will be discussed much more explicitly in the
context of the specific model to be presented in the following Section.

What about the choice of the string scale $\MM_S$, the radius $R$ of the
fifth dimension and the radius $\r$ of the other five (or six)
dimensions that have already been compactified to obtain the
five-dimensional model? In models without any explicit bulk masses
these three parameters are basically fixed if one of the oscillation
phenomena is to be explained by bulk states because one has to obtain
the correct value of the low-energy Planck mass as well as
appropriate masses and mixing of bulk states. In the presence of
bulk masses, however, things are less constrained. For example,
as we have seen from the above discussion, the value of $1/R$ is
not fixed and merely determines how many bulk states actively
contribute to the oscillation. Thus, the phenomenological
possibilities are much richer when bulk masses are allowed.

\section{ Supernova 1987 a}

\label{sec:SN}

One may obtain significant bounds on the mixing angle of a sterile
neutrino from the condition that sterile neutrino emission should not
cool the supernova too much, that is, sterile neutrinos should not carry
away more energy than is carried by doublet neutrinos, approximately
$10^{53}$ ergs. Production of sterile neutrinos proceeds by either
incoherent processes or through coherent oscillation of the active
neutrino state into the sterile Kaluza Klein tower. We shall discuss
both these processes. As we shall see the latter places the most
stringent bounds on the properties of the sterile neutrino.

As a preliminary let us summarise the salient points
\cite{reviews}. The core of the supernova of approximately $10$ km
radius reaches nuclear densities with central density $\rho
_{0}\approx 3.10^{14}\; {\rm g} \,{\rm cm}^{-3}.$ The shockwave moving
into the core produces a hot gas of neutrons, protons, leptons and
radiation. Due to the high density neutrinos are trapped within the
core and following trapping and the establishment of chemical
equilibrium, the lepton fraction, $Y_{L},$ is given by $Y_{L}\sim
0.35.$ The relative contribution to $Y_{L}$ of $e$ and $\nu _{e}$ is
determined by the equilibrium condition on the chemical
potential. Initially the lepton fractions are constant throughout the
core with $Y_{e}=0.28$ and $Y_{\nu _{e}}=0.07,$ corresponding to Fermi
energies $\mu _{e}\sim 235\,\MeV$ and $\mu _{\nu _{e}}\sim 180\,\MeV$.
Thermal distributions of electron neutrinos are maintained by the
dominant charge exchange process
\begin{equation}
e\, p\leftrightarrow \nu _{e}\, n\; .  \label{thermal0}
\end{equation}
The remaining species of doublet neutrinos are produced by the processes 
\cite{raffelt1,thompson} 
\begin{eqnarray}
e^{+}e^{-} &\longrightarrow &\nu \overline{\nu }  \label{thermal} \\
nn &\longrightarrow &nn\nu \overline{\nu }\; .  \notag
\end{eqnarray}
We will assume that the mixing angle of the sterile neutrino with the active
neutrino is small enough that the sterile neutrino can escape from the core.
The case of larger mixing angles has been studied in References \cite
{dolgov,kainulainen} and leads to a lower bound $>0.02$ for the square of
the mixing angle in the medium to the part of the KK tower with the
appropriate mass difference. Apart from an isolated state which may be near
resonance, this corresponds to much larger mixing angle in vacua,
inconsistent with universality constraints at least for the electron and
muon neutrinos~\cite{Ioannisian:99a}.

In calculating the sterile neutrino production rates we need to
distinguish between $\nu _{\mu }$ and $\nu _{\tau }$ production for
which the chemical potential is small \footnote{As discussed in
Ref.~\cite{raffelt} this may be too strong an assumption.} and $\nu
_{e}$ which has a large chemical potential.

\subsection{Incoherent production from $\protect\nu _{\protect\mu }$ and $%
\protect\nu _{\protect\tau }.$}

We first consider the incoherent production of sterile neutrinos from
$\mu $ and $\tau $ neutrinos~\cite{dolgov,BCS}. Incoherent production
takes place if the SM neutrinos are Dirac, with the right-handed
component being sterile. We note that in the model discussed in the
next Section, the SM neutrinos are on the contrary mainly
Majorana-type and so the bound is much weaker. Since the energy stored
in the thermal distribution of $\mu $ and $\tau $ neutrinos is a very
small proportion of the overall energy in the core it is important to
first determine the processes capable of transferring energy from the
electron and nucleon sector to the neutrino sector. The dominant
production processes are those in \eq{thermal}. It turns out that
these processes give similar rates so here we will content ourselves
with a simplified discussion assuming production via the electron
positron channel. The electrons and positrons are in thermal
equilibrium with a gamma radiation which in turn is in equilibrium
with the hot matter behind the shock, that is, with the
neutrons. These processes are much faster than weak and thus we can
consider them as instantaneous, maintaining a thermal distribution of
electrons and positrons. Following reference \cite{dicus} the $\nu $
rate of energy loss is given approximately by
\begin{equation}
Q_{e}=\frac{4}{5}\frac{G^{2}}{\pi ^{3}}((1-C_{V})^{2}+C_{A}^{2})m^{6}\left( 
\frac{\mu _{e}}{m}\right) ^{2}\left( \frac{kT}{m}\right) ^{4}n_{e}e^{-\mu
_{e}/kT}  \label{prodrate}
\end{equation}
where $G$ is the Fermi constant, $m$ is the electron mass,
$n_{f}=N_{f}-N_{\overline{f}}$ where $N_{f}$ is the number density
of fermions and $(1-C_{V})^{2}+C_{A}^{2}\approx 1/4.$ Clearly
$Q_{e}$ is strongly peaked at high temperatures so neutrino production
is dominated by the high initial temperature produced by the shock
wave, $T\approx 30\,\MeV$. To obtain a useful bound it is necessary to
discuss what happens to the neutrinos produced in this process. The
production cross section $Q_{e}$ involves electrons with energy of
$O(\mu _{e})$ while the positron energy integral is approximately
proportional to$\int_{0}^{\infty }dE$ $E^{3}$ $e^{-E/T}.$ One may see
that it predominantly comes from the energy range $E>3T$. Thus the
produced neutrinos have energy much greater than the average
temperature. There are two processes relevant to the fate of these
energetic neutrinos

\begin{itemize}
\item  Elastic scattering from the neutrons. This process causes the
neutrinos to change direction but does not significantly change their
energy. It is the dominant process giving rise to neutrino trapping but not
thermalisation. It also gives the most efficient process for incoherent
production of sterile neutrinos. Following references \cite{BCS,dolgov},
this happens at the rate $\left( m/E\right) ^{2}\Gamma _{nnc}$ where $\Gamma
_{nnc}$ is the neutral, current collision rate on free nucleons 
\begin{equation}
\Gamma _{nnc}\simeq \frac{1}{\pi }G_{F}^{2}N_{B}E_{\nu }^{2}\, .
 \label{scatt}
\end{equation}
and $m$ is the brane bulk mass, as defined in \eq{branebulk}.

\item  Elastic neutral current scattering processes of the neutrinos off
electrons, positrons and neutrinos or neutrino annihilation processes. The
most rapid process is the scattering off electrons because of their greater
abundance due to the large chemical potential. It will rapidly thermalise
the neutrino, reducing the dominant elastic scattering of \eq{scatt}.
However the neutrino remains until it annihilates and during this time the
production of sterile neutrinos continues. The rates for annihilation is
given by 
\begin{equation}
\overline{\Gamma }_{ann}\simeq \frac{G_{F}^{2}}{\pi }N_{\nu }(T)E_{\nu }^{2}
\label{therm}
\end{equation}
so 
\begin{equation}
\frac{\Gamma }{\overline{\Gamma }}\equiv \frac{\Gamma
_{nnc}}{\overline{\Gamma }_{ann}}=\frac{N_{B}}{N_{\nu }}\simeq
3.5\cdot10^{3}\left(\frac{30\,\MeV}{T}\right)^{3}\; .
\label{enhancement}
\end{equation}
\end{itemize}

Then, for small $(m/E),$ the probability the neutrinos oscillate into
sterile neutrinos is simply given by $P=\frac{\Gamma }{\overline{\Gamma }}%
\left( \frac{m}{E}\right) ^{2}RE$ where the factor $RE$ corresponds to the
number of Kaluza Klein states energetically available. If we require that
the energy loss $Q_{e}VtP$ should be less than the energy carried off by
doublet neutrinos we immediately obtain the bound 
\begin{equation}
\frac{m^{2}R}{\eV}<3\cdot 10^{53}\text{ erg }\frac{\overline{\Gamma }}{\Gamma
}\frac{E/eV}{Q_{e}Vt}\simeq 5\cdot 10^{3}e^{\mu _{e}(1/kT-1/k\text{
}30\,\MeV)}
\label{rate}
\end{equation}
where we have assumed a constant density of neutrons and chemical
potential of electron neutrinos ($\mu _{e}\simeq 200\,\MeV)$ over the
volume $V\varpropto r_{\rm core}^{3}\simeq (10^{6}{\rm cm})^{3}$ $.$ Since the
temperature in the supernova falls rapidly over the first ten seconds
and the production cross section is very sensitive to the temperature,
the dominant production occurs within the first second, that is for
$t=1$. In \eq{rate}, $T$ refers to the initial temperature
after the shock and assumes it lasts for one second and we have taken
$E=3T$.

\subsection{Coherent production of $\protect\nu _{\protect\mu }$ and $%
\protect\nu _{\protect\tau }.$}

A much more stringent bound follows from coherent oscillations of the active
neutrino state into the Kaluza Klein tower of sterile neutrino states. \
There are two contributions that must be considered, transitions induced
through level crossing and mixing-induced transitions. The former have
been discussed in references \cite{BCS,Dvali:99a,LRRR}. \ Here we extend
this discussion taking account of the non-vanishing chemical potentials for
the neutrinos and their radial dependence. As we shall discuss there is an
important limiting effect that reduces the effect of level-crossing
conversion. The latter mixing-induced transitions are also much more
important than the incoherent processes considered above. To see this note
that the neutrino flavour state is a coherent mixture of active and sterile
states with an average probability of finding the original flavour state $%
\nu _{a}$ in the $\nu _{s}$ sterile state given by $\sin ^{2}(2\theta )/2$
where $\theta $ is the $\nu _{a}$, $\nu _{s}$ mixing angle. The
mixing-induced transition rate is given by $\Gamma _{nnc}$ $\sin
^{2}(2\theta )/2$ to be compared with the incoherent production rate of $%
\Gamma _{nnc}(m/E)^{2}.$ The mixing angle to the sterile neutrino tower in
vacua is approximately $m/M_{n}$, much larger than the equivalent factor $%
(m/E)$ relevant to incoherent production. In the medium of the supernova the
mixing can be much larger due to resonant effects because the electron
doublet neutrino can be degenerate with one of the KK levels. Thus the
mixing-induced transitions will also be much larger than the
non-resonant transitions and can compete with the level crossing
contributions. Here we will compute only the level-crossing contribution; a
detailed calculation of the full contribution will be presented elsewhere.

Let us consider the muon neutrino case - an identical discussion applies to
the tau neutrino. The muon neutrino effective potential, $V_{\mu },$ in the
supernova core is given by 
\begin{equation}
V_{\mu }=\sqrt{2}G(2n_{\nu _{\mu }}+n_{\nu _{e}}+
         n_{\nu _{\tau }}-0.5n_{n})\; .
\label{effpot}
\end{equation}
Initially only $n_{\nu _{e}}$ and $n_{n}$ are significant yielding a
negative $V_{\mu }$, of order $-10\,\eV$. In this case it is the
antineutrino, $\overline{\nu }_{\mu },$ that crosses the Kaluza-Klein
levels through matter effects. It acquires a resonant mass
$m_{n}\simeq n/R\simeq (2V_{\mu }E)^{1/2}\simeq 10^{4.5}\,\eV$. The
effect of these oscillations can be reliably estimated for small
mixing angles, because the width of each resonance is smaller than the
separation between resonances. \ Thus the survival probability for
a standard neutrino produced in the core is given by the product of
the survival probabilities in crossing each resonance $P_{\nu \nu
}\simeq \Pi _{n}P_{n}$, where $P_{n}\simeq e^{-\pi \gamma _{n}/2},$
and
\begin{equation}
\gamma _{n}\simeq \frac{4m^{2}}{E}\frac{V}{dV/dr}\; .  \label{gamma}
\end{equation}
Following reference \cite{BCS} we approximate the density profile by
$\rho (r)\varpropto e^{-r/r_{\rm core}}$. This gives $\
V/(dV/dr)=r_{\rm core}$ and hence $\gamma _{n}\simeq
m^{2}/10^{-3}\,\eV^{2}.$

To determine the energy loss due to coherent processes we must determine the
survival time of the active neutrino after production. \ The neutrinos
initially produced by the annihilation processes \eq{thermal}
thermalise through the elastic scattering $\nu _{\mu }$ $e\rightarrow \nu
_{\mu }$ $e$. The electron abundance is very large due to its large chemical
potential and so thermalisation occurs on a short timescale compared to that
for $\nu \overline{\nu }$ annihilation which stops the level-crossing
production process. Thus most of the production of sterile neutrinos occurs
after the (anti) neutrino acquires its thermal distribution. As the
antineutrino moves towards the surface of the supernova its mass decreases
due to the density change and thus it crosses several of the Kaluza-Klein
levels before it annihilates. To estimate the number of levels crossed we
first compute the mean free path, $\lambda ^{\prime }(T,E_{\nu }),$ before $%
\nu \overline{\nu }$ annihilation. The annihilation cross section is 
\begin{equation}
\sigma _{_{\nu \overline{\nu }\longrightarrow e^{+}e^{-}}}\simeq \frac{%
G_{F}^{2}}{\pi }n_{\nu }(E_{\nu },T)E_{\nu }^{2}
\end{equation}
giving 
\begin{equation}
\lambda ^{\prime }(T,E_{\nu })^{-1}=C\int_{0}^{\infty
}E_{\overline{\nu }}^{2}\,\sigma _{\nu \overline{\nu
}\longrightarrow e^{+}e^{-}}\,(E_{\overline{\nu }}+E_{\nu
})e^{-E_{\overline{\nu }}/T}dE_{\overline{\nu }}\; .
\end{equation}
It is convenient to eliminate the constant $C$ by writing this as 
\begin{eqnarray}
\lambda ^{\prime }(T,E_{\nu })^{-1} &=&\lambda _{0}^{\prime }(T,E_{\nu
}=100\,\MeV )^{-1}\cdot\left( \frac{\lambda ^{\prime }(T,E_{\nu
})^{-1}}{\lambda _{0}^{\prime }(T,E_{\nu }=100\,\MeV )^{-1}}\right)
\label{scatter} \\ &=&\lambda _{0}^{\prime }(T,E_{\nu
}=100\,\MeV )^{-1}\cdot\left( \frac{\int_{0}^{\infty
}E_{\overline{\nu }}^{2}(E_{\overline{\nu }}+E_{\nu
})^{2}e^{-E_{\overline{\nu }}/T}dE_{\overline{\nu }}}{\int_{0}^{\infty
}E_{\overline{\nu }}^{2}\, e^{-E_{\overline{\nu }}/T}dE_{\overline{\nu
}}\left ( 100\,\MeV\right) ^{2}}\right) \notag
\end{eqnarray}
where $\lambda _{0}^{\prime }$ is defined by $(\lambda _{0}^{\prime
})^{-1}=n_{\nu }(T)\sigma _{\nu \overline{\nu }\longrightarrow
e^{+}e^{-}}(E)$, that is,
\begin{equation}
\lambda _{0}^{\prime }(T,E_{\nu }=100\,\MeV )=2\cdot 10^{4}\left(
\frac{30\,\MeV}{T}\right) ^{3} \text{cm}\; .
\end{equation}
Using this we may now determine the energy loss, $E_{S},$ into the sterile
neutrino tower. Averaging over the thermal energy distribution of the
neutrino, we find 
\begin{equation}
E_{S}=\frac{1}{2}\frac{\pi }{2}Q_{e}(T)V\text{ }t\frac{\int_{0}^{\infty }
\text{ }\delta n(E,T)\gamma (E)E^{3}e^{-E/kT}dE}{<E>\int_{0}^{\infty }\text{ 
}E^{2}e^{-E/kT}dE}
\end{equation}
where the first factor of 1/2 comes because only $\overline{\nu }_{\mu }$
resonates and $<E>$ is the mean energy of the neutrinos produced by
annihilation. From \eq{gamma} we have 
\begin{equation}
\gamma (E)=\frac{4m^{2}}{E}\frac{V}{dV/dr}=\left(
\frac{m^{2}}{10^{-3}\,\eV ^{2}} \right) \left(
\frac{100\,\MeV}{E}\right)
\end{equation}
and from \eq{scatt} 
\begin{eqnarray}
\delta n &=&(\delta \rho /2\rho )m_{n}R, \\ \delta \rho /\rho
&=&\frac{\lambda ^{\prime }(T,E_{\nu })}{r_{\rm core}}\simeq
\frac{1}{2}\, 2\cdot 10^{-2}\left( \frac{30\,\MeV}{T}\right)
^{3}\left( \frac{ \int_{0}^{\infty }E_{\overline{\nu
}}^{2}(E_{\overline{\nu }}+E_{\nu })^{2}e^{-E_{\overline{\nu
}}/T}dE_{\overline{\nu }}}{\int_{0}^{\infty }E_{ \overline{\nu
}}^{2}\, e^{-E_{\overline{\nu }}/T}dE_{\overline{\nu }}\left(
100\,\MeV\right) ^{2}}\right) ^{-1} \notag \\ m_{n} &=&2\left (
\mathcal{V}E\right) ^{1/2}\simeq 10^{5}\,\eV\left( \frac{ E_{\nu
}}{100\,\MeV}\right) ^{1/2} \notag
\end{eqnarray}
where the first factor of 1/2 in $\delta \rho $ comes from the fact that the
neutrino often elastically scatters so that it moves in a direction along
which the density does not change. Putting all this together gives 
\begin{eqnarray}
E_{S} &=&\frac{1}{2}\frac{\pi }{2}Q_{e}(T)\text{ }V\text{ }t\text{
}\frac{ m^{2}R}{4}\, 2\cdot 10^{-2}\left( \frac{30\,\MeV}{T}\right)
^{3}\left( \frac{10^{5}\,\eV}{ 10^{-3}\,\eV^{2}}\right) B \\ B
&=&\frac{(100\,\MeV)^{2}}{<E>}\int_{0}^{\infty }\frac{E^{3}e^{-E/T}}{
\int_{0}^{\infty }E_{\overline{\nu }}^{2}(E_{\overline{\nu
}}+E)^{2}e^{-E_{ \overline{\nu }}/T}dE_{\nu }}\left(
\frac{100\,\MeV}{E}\right) ^{1/2}dE \notag
\end{eqnarray}
and hence
\begin{equation}
E_{S}=10^{60}\left( \frac{T}{30\,\MeV}\right) e^{-\mu _{e}(1/kT-1/k\text{ }
30\,\MeV)}\frac{m^{2}R}{eV}\,B\; .
\end{equation}
Evaluating the integrals numerically results in
\begin{equation}
B\simeq 0.5\left( \frac{30\,\MeV}{T}\right) ^{2.5}
\end{equation}
so finally 
\begin{equation}
E_{S}\simeq 4\cdot 10^{61}\left( \frac{30\,\MeV}{T}\right) ^{1.5}e^{-\mu
_{e}(1/kT-1/k\; 30\,\MeV)}\frac{m^{2}R}{\eV}\; .
\end{equation}
Requiring that this be bounded by $3\cdot 10^{53}$ ergs implies 
\begin{equation}
\frac{m^{2}R}{\eV}<6\cdot 10^{-7}\left( \frac{T}{30\,\MeV}\right)
^{1.5}e^{\mu _{e}(1/kT-1/k\; 30\,\MeV)}\; . \label{bound1}
\end{equation}

\subsubsection{Self-limiting of level-crossing processes}

In fact this bound is too strong because in deriving it we assumed
$n_{\nu _{\mu }}$ was negligible. If the bound is violated then, on a
timescale much less than one second, a significant number of muon
anti-neutrinos oscillate into sterile neutrinos. However the muon
neutrinos remain as they do not undergo resonant conversion provided
$V_{\mu }$ is negative. Thus a non-zero value of $n_{\nu _{\mu }}$
builds up driving $V_{\mu }$ to zero and thus stopping the resonant
conversion\footnote{ A similar effect has been considered in
Ref.~\cite{Nunokawa:1997ct}.}. One may readily determine the energy
loss before this happens because the change $ \delta n_{\nu _{\mu }}$
needed is $n_{n}/4-n_{\nu _{e}}/2=0.13\, N_{B}.$ Since the muon
neutrinos are pair produced, an equal number of muon antineutrinos
must have converted into sterile neutrinos. However since the
antineutrinos are not degenerate and, as discussed above, they have a
thermal distribution before resonant conversion, they carry
approximately $3T$ energy. Thus the energy loss is approximately
$1/7$th of the energy stored in the electron and neutrinos sea in the
core of the supernova, probably just acceptable.

Does this mean there is no constraint on the muon (and tau) mixing
parameter $m?$ This depends on whether the mechanism setting $V_{\mu
}$ to zero is completely efficient. In the first half second half of
the leptons leave the core \cite{latimer}. If \eq{bound1} is not
satisfied this reduction will be made up by a further $\delta ^{\prime
}n_{\nu_{\mu }} =\delta n_{\nu _{e}}/2\approx n_{\nu _{e}}/4=0.02\,
n_{B}$, a relatively minor correction. More important is the fact that
diffusion processes might spoil the local cancellation of $V_{\mu
}$. A crude estimate of this effect may be obtained by noting that
the neutrino cannot be localised within its mean free path,
$\lambda_{\rm mfp}$. As a result, even though when averaged over a mean
free path the average value, $<V_{\mu }>$, is zero, it may not vanish
locally. Thus one may expect resonant conversion of $\nu _{\mu }$ to
occur locally in regions in which $V_{\mu }$ is positive and
conversion of $\overline{\nu }_{\mu }$ in regions in which $V_{\mu }$
is negative. Together these will not change the lepton number and
hence leave $<V_\mu >$ zero. A naive estimate of the sterile neutrino
production due to this effect may be obtained if one notes that over
the mean free path the nucleon density changes by an amount given by
$\delta n_{n}/n_{n}=\lambda _{\rm mfp}/r_{\rm core}.$ Hence locally \
$\delta V_{\mu }\approx \sqrt{2}Gn_{n}\lambda _{\rm mfp}/2r_{\rm core}.$
The overall energy loss is proportional to $\sqrt{\delta V_{\mu }}$
and hence the rate for resonant conversion when $<V_{\mu }>=0$ is
reduced by the factor $\sqrt{\lambda _{\rm mfp}/r_{\rm core}}$ compared to
that given in \eq{rate}. The mean free path is determined by the
elastic scattering off nucleons via \eq{scatt}. Using it we find the
revised bound
\begin{equation}
\frac{m^{2}R}{\eV}<10^{-5}\left( \frac{T}{30\,\MeV}\right) ^{0.5}e^{\mu
_{e}(1/kT-1/k\text{ }30\,\MeV)}  \label{bound2}
\end{equation}
This applies to the brane bulk mixing of both muon and tau neutrinos. Notice
that if the overall neutrino distribution were able to locally compensate
the variation of the neutron density the bound would become weaker. Analysis
of this possibility requires constructing a generalised diffusion equation -
work on this is in progress.

\subsection{Incoherent production from $\protect\nu _{e}$}

The case of $\nu _{e}$ is somewhat different from $\nu _{\mu }$ and
$\nu _{\tau }$ because the $\nu _{e}$ pathlength before it is
eliminated by the process $\nu _{e}+n$ $\rightarrow $ $e+p$ is much
shorter than the pathlength for $\nu \overline{\nu }$ annihilation
which is relevant to the case of $\nu _{\mu }$ and $\nu _{\tau }.$ As
a result the $\nu _{e}$ elimination pathlength is approximately the
same as the elastic scattering pathlength.  Hence, there is no
enhancement factor corresponding to \eq{enhancement} and the bound is
given by \eq{rate} with $\bar{\Gamma}/\Gamma=1$, leading to a bound
some three orders of magnitude weaker.

\subsection{Coherent production from $\protect\nu _{e}$}

The analysis of the case of resonant conversion follows the general
lines discussed above for the mu and tau neutrinos. In this case 
\begin{equation}
V_{e}=\sqrt{2}G(2n_{\nu _{e}}+n_{e}+n_{\nu _{\mu }}+n_{\nu _{\tau
}}-0.5n_{n}).
\end{equation}
An immediate difference is that $V_{e}$ is positive and somewhat
smaller, namely $V_e\approx 3\,\eV.$ This means that it is the $\nu _{e}$
that undergoes resonant conversion and the process stops when $V_{e}$
is zero. This corresponds to $ \delta n_{\nu _{e}}=-(2n_{\nu
_{e}}+n_{e}-0.5n_{n})/2=-0.06\, n_{B}$ corresponding to only $1/16$th of
the total energy. Subsequently diffusion causes half of the lepton
number to leave the core in the first half second and to bring $V_{e}$
back to zero requires resonant conversion of $\overline{ \nu }_{e}$
with $\delta ^{\prime }n_{\nu _{e}}=n_{n}/8=-0.09\, n_{B}$ giving a total
loss slightly more than $1/8$th of the total energy. Given that
diffusion continues for several seconds the losses may become
unacceptably large, but a detailed simulation is necessary to resolve
this question.  Assuming that the loss is acceptable one must finally
consider the losses due to the fact $V_e$ is not locally zero. To
determine this we note that the relevant production process is
$e^{+}e^{-}\longrightarrow \nu _{e} \overline{\nu }_{e}$ rather than
the more rapid process $e+p\rightarrow \nu _{e}+n.$ As discussed
above, conversion occurring with $<V_{e}>$ zero conserves lepton
number and adjusts itself so that the conversion rate for $\nu _{e}$
and $\overline{\nu }_{e}$ are equal. \ Thus it is sufficient to
determine the conversion rate for $\overline{\nu }_{e}.$ The energy of
the $\overline{\nu }_{e}$ is only slightly larger than $3T$ and we
will take it to be $100\,\MeV$ in our estimates. The annihilation rate is
dominated by the process $\overline{\nu }_{e}+p\rightarrow e^{+}+n$
and the corresponding annihilation pathlength is $\lambda ^{\prime
}(T,E_{\overline{\nu } _{e}})=30\left( \frac{100\,\MeV}{E_{\nu }}\right)
^{2}$ cm. The resulting energy loss is given by
\begin{equation}
\frac{dE_{S}}{dE_{\nu }}=\frac{1}{2}\frac{\pi }{2}\frac{dQ_{e}(T,E_{\nu })}{
dE_{\nu }}V\text{ }t\text{ }\delta n\gamma
\end{equation}
where $\int \frac{dQ_{e}(T,E)}{dE}dE=$ $Q_{e}(T)$ is given by
\eq{prodrate}, $\delta n=\frac{\lambda ^{\prime }(T,E_{\nu })}{r_{\rm
core}} m_{n}R, $ and $\gamma $ is taken from \eq{gamma}. We
estimate $m_{n}$ for the case $<V_{e}>=0$ in the manner described
above giving $ m_{n}=\sqrt{\lambda _{\rm mfp}/r_{\rm core}}\, 10^{5}\left(
\frac{E_{\nu }}{100\,\MeV} \right) ^{1/2}\eV$. Since the neutrino chemical
potential is much less than that of the electron the energy of the
electron and positron will roughly be shared between the neutrino and
the antineutrino. For the calculation of $\delta n$ and $\gamma$ we
use an average neutrino energy of $150\,\MeV$.  Using this to evaluate
$E_{S}$ leads to the bound
\begin{equation}
\frac{m^{2}R}{\eV}<3\cdot 10^{-1}\left( \frac{T}{30\,\MeV}\right)
^{0.5}e^{\mu _{e}(1/kT-1/k\; 30\,\MeV)}
\end{equation}

\section{An explicit orbifold model}

\label{sec:5}

\noindent

In this section, we consider a model with small mixing between SM and
bulk neutrinos. As a consequence, this mixing can be treated
perturbatively and it is unlikely that bulk neutrinos play a relevant
role in the conversion of atmospheric $\nu_\mu$. On the other hand, a
small bulk component in the electron neutrino might be responsible for
the depletion of solar neutrinos. Such a possibility, first considered
in Ref.~\cite{Dvali:99a}, has been investigated in Ref.~\cite{LRRR} in
the context of a model with Dirac bulk masses~\cite{LR}. A detailed
phenomenological analysis has shown that one may go from the limit in
which only a single bulk state is involved in the resonant conversion
to the case where a continuum contributes. The former gives the same
phenomenology as the standard single sterile neutrino case and is
disfavoured when the SuperKamiokande (SK) data on the recoil electron
spectrum is taken into account. However, if more bulk states
contribute, the deviation from the observed spectrum reduces and a
very good fit of the data is possible for a level spacing in which
three KK modes contribute significantly to the resonant
oscillation. In~\cite{LRRR} it has also been shown that the model can
be extended to describe all present indications of neutrino masses,
including maximal mixing for atmospheric neutrinos in a natural way
and the LSND signal. In this section, we will show that the same
phenomenology can be reproduced in the context of a $\ZZ$ orbifold
model by means of Majorana-type bulk mass terms.

\subsection{General structure of the model}

\label{sec:mm}

\noindent

We consider a $\ZZ$ orbifold model or
equivalently, a model with $\ZZ$ invariance. The $\ZZ$ symmetry is
generated by the action of $P_5$ defined in \eq{P5}. 
It can be seen from Table~\ref{table:symmetries}, that only Majorana
bulk masses are allowed by the $\ZZ$ symmetry. Also, only the
brane-bulk mixing terms $\mbb$, connecting the SM neutrinos to the
$\xi$ components of the fermion fields, are allowed while $\mbbc =
0$. In addition, we assume that the vector-like Majorana term
vanishes, that is $\MM_V = 0$. This follows, if the mass Lagrangian
respects the underlying five-dimensional Lorentz
invariance. Alternatively, vector-like Majorana masses generated by the
VEV of a 5-dimensional vector field are forbidden if one assumes that
the representation of the gauge group does not mix $\Psi$ and
$\Psi^c$. Furthermore, for ease of notation, we write $M\equiv M_S$ for
the Majorana mass.

Since $\MM$ is the only bulk mass matrix, it is convenient to consider a
basis for the bulk fields in which this matrix is diagonal, real
and positive, that is $\MM = \diag(\MM_1\ldots \MM_N)$. We continue to
call the brane-bulk mixing in this basis $\mbb$. Specializing the
general expression~\eqref{mp} to our case, one can then explicitly sum
the series to obtain the mass matrix
\begin{equation}
\label{mlight}
\ml_{ij} = -\sum_{I,n\in\Z}\frac{\mbb_{Ii}\mbb_{Ij}\MM_I}{\MM^2_I+n^2/R^2}
 = -\sum_{I}\pi R \mbb_{Ii}\mbb_{Ij} \coth (\pi R \MM_I) \; .
\end{equation}
This equation clearly shows a see-saw type structure, where the
``effective'' heavy mass matrix $M_H = \diag(M^H_1\ldots M^H_N)$
is specified by 
\begin{equation}
\label{mright}
M^H_I = \left(\pi R\coth(\pi R \MM_I)\right)^{-1} \; .
\end{equation}
The heavy masses $M^H_I$ involved in the see-saw formula~(\ref{mlight})
is roughly speaking the lighter one of the masses $\MM_I$ and $(\pi
R)^{-1}$. In the limit $\MM_I\ll (\pi R)^{-1}$ we have, in fact,
$M^H_I\simeq\MM_I$, whereas the opposite limit $(\pi
R)^{-1}\ll\MM_I$ leads to $M^H_I\simeq (\pi R)^{-1}$. These results can
be easily understood. In the first case, only the
lowest bulk modes with $n=0$ and mass $\MM_S$ effectively contribute
while all states with $n\neq 0$ are much heavier.  Correspondingly, we
find an ``ordinary'' see-saw mechanism with the $n=0$ bulk modes as
heavy neutrinos and corresponding heavy masses $\MM_I$. In the
opposite case, when $(\pi R)^{-1}\ll\MM_I$, the series~\eqref{mlight}
is dominated by terms satisfying $n^2/R^2\lesssim \MM^2_I$ which are of order
$-\mbb_{Ii}\mbb_{Ij}/\MM_I$. Since the number of those terms is
$\ord{\MM_I R}$, $\MM_I$ cancels. It is replaced by
$R^{-1}$ resulting in a ``geometrical'' see-saw
mechanism~\cite{Dienes:98a}.

The mass eigenstates follow from diagonalization of the mass
Lagrangian, \eq{Lmassp}. Actually, as far as neutrino oscillations are
concerned, we only need to diagonalize the hermitian matrices
${\ml}^\dagger \ml$ and $\cM_n^\dagger \cM_n$. Using
$\cM_n^\dagger\cM_n$ is clearly more convenient in the present case
than using $\cM_n$. In fact, since $M$ is the only non-zero mass
term, $\cM_n^\dagger\cM_n$ is diagonal, as can be easily seen from
eq.~\eqref{Mn}. This implies that the fields $\eta'_n$ and $\xi'_n$ in
\eqs{basis}, although not being mass eigenstates of $\cM_n$, are
indeed eigenstates of $\cM_n^\dagger\cM_n$. Evaluating
eq.~(\ref{nueigen}), the electroweak eigenstates as a superposition of
mass eigenstates are given by
\begin{equation}
\label{flavour1}
\nu_i = \sum_{k=1,2,3}U_{ik}\hat\nu_k + \sum_{I,n\in\Z}\frac{
\mbb_{Ii}^*}{\sqrt{\MM^2_I+n^2/R^2}} \; \frac{\MM_I \, \xi'_{nI} +in/R \,
\eta'_{nI}}{\sqrt{\MM^2_I+n^2/R^2}} \; .
\end{equation}
Here $\hat\nu_k$ are the light mass eigenstates of mass $m_k$. They
are defined by $\nu_k' = U_{ki}\hat\nu_i$ where the unitary matrix $U$
diagonalizes ${\ml}^\dagger\ml$, that is,
$\diag(m_1,m_2,m_3)=U^\dagger{\ml}^\dagger \ml U$.

For each $n>0$, we can define the $\ZZ$ eigenstates
\begin{equation}
\xi_I^\pm = \frac{\xi'_{nI} \pm \xi'_{-nI}}{\sqrt{2}}\; , \qquad
\eta_I^\pm = \frac{\eta'_{nI} \pm \eta'_{-nI}}{\sqrt{2}} \; ,
\end{equation}
where $\xi_I^+$, $\eta_I^-$ are even and $\xi_I^-$, $\eta_I^+$ are odd.
By combining modes with opposite $n$ in \eq{flavour1}, we see that the
two $\ZZ$-odd combinations do not mix with the SM neutrinos. This was
expected from the general discussion at the end of Section 2. Moreover,
only the $\ZZ$-even combination
\begin{equation}
\label{coupled}
\nuh_{nI} \equiv \frac{\MM_I \, \xi^+_{nI} +in/R \,
\eta^-_{nI}}{\sqrt{\MM^2_I+n^2/R^2}} \qquad n\geq 1
\end{equation}
does mix with the SM neutrinos. This means that only one out of the
four degenerate states $\xi'_{nI}$, $\eta'_{nI}$, $\xi'_{-nI}$,
$\eta'_{-nI}$ actually couples to the SM neutrinos. The mass
$M_{nI}$ of all these states is given by~\footnote{In
Ref.~\cite{Dienes:98a} an ``effective'' Majorana mass term was
considered that was induced by a Scherk-Schwarz type compactification.
In the context of our models with explicit bulk mass terms such a
spectrum can be obtained by vector-like Majorana bulk masses and
all other mass terms vanishing. This can be explicitly seen from
\eq{Mspec}.} 
\begin{equation}
\label{massesZ2}
M_{nI} = \sqrt{\MM_I^2 +\frac{n^2}{R^2}} \; .
\end{equation}

Analogously, in the zero-mode sector the $\ZZ$-even field
\begin{equation}
\nuh_{0I}\equiv \xi'_{0I} \; ,
\end{equation}
with mass $M_{0I} = \MM_I$ mixes with the SM neutrinos, whereas
$\eta'_0$ does not.  We can, therefore, rewrite \eq{flavour1} as follows
\begin{equation}
\label{flavour}
\nu_i = \sum_{k=1,2,3}U_{ik}\hat\nu_k + \sum_I\frac{
\mbb_{Ii}^*}{\MM_I}\, \nuh_{0I} + \sqrt{2}
\sum_{I,n>0}\frac{
\mbb_{Ii}^*}{\sqrt{\MM^2_I+n^2/R^2}} \, \nuh_{nI} \; .
\end{equation}
\Eq{flavour} expresses the SM flavour eigenstates $\nu_e$, $\nu_\mu$,
$\nu_\tau$ in terms of the light eigenstates $\hat\nu_k$,
$k=1,2,3$ and the tower of heavy eigenstates $\nuh_{nI}$, $n\geq 0$ and
is, therefore, the starting point for studying the oscillation
phenomenology in the perturbative regime. The exact formulae with
matter effects included are explicitly presented in
Appendix~\ref{app:diag} for the case of one family of SM and bulk
fields.

\medskip

Had we considered a U(1)-symmetric model with Lorentz-invariant Dirac
bulk mass term only~\cite{LRRR}, we would have found similar results.
The flavour eigenstates would still be given by \eq{flavour} in terms
of light eigenstates $\hat\nu_k$ and a tower of mass eigenstates
$\nuh_{nI}$, where $n\geq 0$. However, those states are now
expressed in terms of the original states $\eta_{nI}$ and $\xi_{nI}$
as
\begin{equation}
\nuh_{0I} \simeq \eta_{0I} \qquad \nuh_{nI} \simeq
\frac{e^{-i\phi_{nI}}\eta_{nI} + e^{i\phi_{nI}} \eta_{-nI}}{\sqrt{2}}
\; .
\end{equation}
where the angles $\phi_{nI}$ are defined by the relation $\MD_I+in/R =
e^{i\phi_{nI}}\sqrt{\MD_I^2+n^2/R^2}$.  The masses of the $n$-th mode
is given by \eq{massesZ2} but with $\MM_I$ replaced by the eigenvalues
$\MD_I$ of the Dirac mass matrix. A crucial difference is that the
light states $\hat\nu$ are massless for an unbroken U(1) symmetry. In
a general context, we have already observed this property in the
previous Section. The mixing matrix in \eq{flavour}, therefore,
acquires physical meaning only once the U(1) symmetry is broken, for
example due to small U(1) breaking brane-bulk masses or to small bulk
Majorana masses.
\medskip

Returning to the $\ZZ$ model, we should analyze the validity of the
perturbative diagonalization that led to eq.~\eqref{flavour}.
Applying the discussion of Section~\ref{bbcouplings}, in particular
eq.~\eqref{e2}, we find that
\begin{equation}
\label{pertcon}
\e^2 = \sum_I\frac{\pi R |\mbb_{Ii}|^2}{\MM_I}\coth(\pi R\MM_I) \; ,
\end{equation}
should be smaller than one in order to be in the perturbative regime.
Then the heavy eigenstate components of each SM neutrino
flavour $i = e,\mu,\tau$ in \eq{flavour} are small. Therefore, the
oscillations of SM neutrinos mainly involve the three light Majorana
states $\hat\nu_k$, $k=1,2,3$ and the corresponding mixing matrix $U$
can be identified with the MNS matrix. As discussed above, for such models,
some of the observed oscillation phenomena can be attributed to oscillations
into the three light Majorana neutrinos (playing the role of the
three standard neutrinos) while some others can be due to oscillation
into bulk states.

Since we consider the case that the sterile states play a relevant
role in solar oscillations only all other oscillation phenomena
should be due to oscillation between the light Majorana states. We
will see that we can naturally obtain $\nu_\mu\leftrightarrow\nu_\tau$
oscillation with maximal mixing to account for the atmospheric
neutrino deficit. At the same time, we obtain
$\nu_e\leftrightarrow\nu_\mu$ oscillations with a $\dm{}$ larger than
the atmospheric one, that can be used to accommodate the LSND signal,
and two mainly SM neutrinos with masses in the eV range, that can
provide a cosmologically significant dark matter component.  We start
with an analysis of solar neutrinos in the simple one-family case
which we will, subsequently, embed in the full three-family model.

\subsection{Solar neutrinos}

\label{sec:solar}
\noindent

We first assume that $\nu_e$ only mixes with a single bulk fermion
$\Psi_e$ with mass $\MM_e$ through a brane-bulk mass term
$\mbb_e$. Specializing eq.~\eqref{flavour}, the electron neutrino can
be written as the superposition
\begin{equation}
\label{flavour1f}
\nu_e \simeq \hat\nu_e + \frac{\mbb_e}{\MM_e}\, \nuh_{0} + \sqrt{2}
\sum_{n=1}^\infty\frac{
\mbb_{e}}{\sqrt{\MM^2_e+n^2/R^2}} \, \nuh_{n} \; ,
\end{equation}
where $\mbb$ has been taken positive without loss of generality.
The states $\nuh_n$, where $n\geq 0$, have mass
$M_n\simeq\sqrt{\MM^2_e + n^2/R^2}$, whereas the mass of the light
state $\hat\nu_e$ is given by
\begin{equation}
\label{me}
\ml_{e} \simeq \pi R \mbb_e^2\coth(\pi R \MM_e)\; .
\end{equation}
Notice that the condition $\e^2\ll 1$ characterizing the perturbative
regime is equivalent to $\ml_{e}\ll\MM_e$ in this case.

The corrections to \eq{flavour1f} and to the mass eigenvalues, as well
as the behaviour in the presence of matter effects can be obtained by
using the exact equations in Appendix~\ref{app:majorana}. For
completeness, in Appendix~\ref{app:dirac}, we have also given the
exact equations for the model with U(1) lepton number symmetry as
briefly discussed in the previous Subsection. From \eqs{sysunpert} we
see that the $\ZZ$-odd states $\xi^-_n$ for $n\geq1$ and $\eta^+_n$
for $n\geq0$, are decoupled as expected, a consequence
of the exact $\ZZ$ parity symmetry. Moreover, from an analysis of
\eq{majoval}, the structure of the mass spectrum in presence of matter
effects turns out to be the following.
\begin{itemize}
\item
There exists one ``light'' Majorana eigenstate $\hat\nu$ with mass
$\lambda '$ in the range $0 < \lambda '< \MM_e$. In vacuum, $\lambda '$ is
approximately  given by the see-saw formula~(\ref{me}) implying that
$\lambda '\simeq {m '}_e\ll\MM_e$.
\item
We have one ``zero-mode'' eigenstate $\nuh_0$ with squared mass
$\lambda^2_0$ in the range $\MM^2_e < \lambda^2_0 < \MM^2_e
+(1/R)^2$. In vacuum,
\begin{equation}
\lambda^2_0 \simeq \MM^2_e+2\mbb_e^2 \simeq \MM^2_e\; .
\end{equation}
\item
For each $n\geq 1$ there are two modes $\nuh^\pm_n$ with masses
$\l_{n\pm}^2$ in the range $M_e^2+(n/R)^2<\l_{n\pm}^2<M_e^2+((n+1)/R)^2$.
Their masses in vacuum are given by 
\begin{equation}
\lambda^2_{n\pm} \simeq \MM_e^2+\frac{n^2}{R^2}+2\frac{\mbb_e^2}{n}
\frac{\sqrt{\MM_e^2+ n^2/R^2}\pm\MM_e}{\sqrt{\MM_e^2+ n^2/R^2}}
\simeq \MM_e^2+\frac{n^2}{R^2}
\; .
\end{equation}
\end{itemize}

\begin{figure}[t]
\begin{center}
\epsfig{file=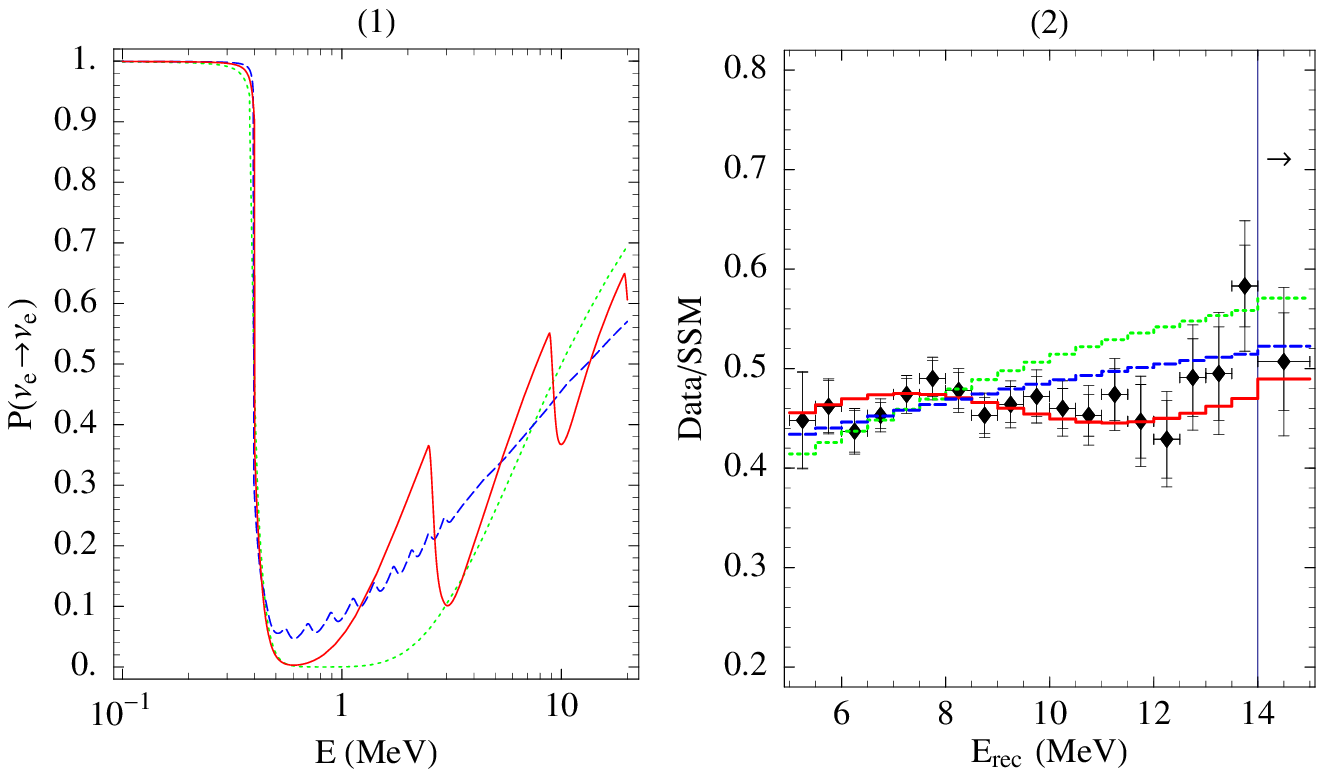,width=1.00\textwidth}
\end{center}
\mycaption{Survival probabilities for $\MM_e\simeq 2.1\cdot
10^{-3}\eV$ and $(1/R,\mbb_e)=(20, 0.058)\cdot 10^{-3}\eV$ (dotted
line), $(1/R,\mbb_e)=(5, 0.035)\cdot 10^{-3}\eV$ (solid line),
and $(1/R,\mbb_e)=(0.5, 0.012)\cdot 10^{-3}\eV$ (dashed line).}
\label{fig:sur} 
\mycaption{Recoil energy spectrum associated with the three survival
probabilities in \Fig{sur}.}
\label{fig:spe} 
\end{figure}

We see from the last equation that the two $n$-modes are almost degenerate
so that their linear combination $\nuh_n$ in \eq{flavour1f} can be
considered as approximate mass eigenstates; the orthogonal
combination is approximately decoupled. The resonance involving the
bulk state
$\nu_n$ becomes relevant for the oscillations of a solar $\nu_e$ if
its energy $E_\nu$ is such that 
\begin{equation}
\label{res}
A_e\equiv 2 E_\nu V_e \simeq M^2_{n} \; .
\end{equation}
Here $V_e$ is the matter induced potential given in terms of the
electron and neutron densities $n_e$, $n_n$ by $V_e = \sqrt{2}\, G_F
(n_e - n_n/2)$.  The energy-width $(\delta E)_n$ of the $n$-th
resonance, namely the width of the energy range where the mixing
$\theta_n$ with the $n$-th bulk states is large, can be precisely
defined by the condition $\sin^22\theta_n>1/2$. It then follows that
$(\delta E)_n\simeq 2\sqrt{2} \mbb_e/M_n E_n$, where $E_n$ is the
resonant energy. Notice that, when expressed in terms of the parameter
$\sqrt{A_e}$, the width is independent of the resonance, that is
$(\delta\sqrt{A_e})_n\simeq \sqrt{2} \mbb_e$. Since $(\delta E)_n\ll
E_{n+1}-E_n$, the resonances take place in a small portion of the
energy range. This fact, which has already been used in
Section~\ref{sec:SN}, considerably simplifies the phenomenological
analysis. When the neutrino energy is far from the resonant energy,
the electron neutrino is mainly made of the light mass eigenstate
$\hat\nu$ if $A_e< \MM^2_e$ and it is mainly made of the $n$-th mass
eigenstate $\nuh_n$ if
\begin{equation}
\MM^2_e + \frac{n^2}{R^2} < A_e < \MM^2_e + \frac{(n+1)^2}{R^2} \; .
\end{equation}
This determines the behaviour of the electron neutrino survival
probability.  When the neutrino energy is so small that the value of
$A_e$ in the core is smaller than the first resonance, that is $E_\nu
< \MM^2_e/(2V_e^{\text{core}})$, the neutrinos predominantly consist
of the light mass eigenstate $\hat\n$ and the survival probability is
close to 1. For energies larger than $\MM_e^2/(2V_e^{\text{core}})$,
on the other hand, the produced neutrinos are almost orthogonal to the
detected ones and the survival probability in the adiabatic limit is
small. The energy dependence of the solar neutrino depletion therefore
determines $\MM_e$.  The Majorana mass $\MM_e$ must be in fact in the
range $\MM_e = (2\dash 3)\cdot 10^{-3}\eV$, in such a way that solar
neutrinos in the Beryllium energy range undergo resonant conversion
but $pp$--neutrinos do not. When the neutrino energy increases, the
adiabatic approximation fails, the level crossing probabilities grow
and the survival probability between two subsequent resonances also
grows. The rate of growth between resonances depends on the mixing in
vacuum between the electron neutrinos and the mass eigenstates which,
for given $\MM_e$ and $R^{-1}$ is determined by the brane-bulk mass
$\mbb_e$.  On the other hand, the survival probability falls after
each resonance, since the number of levels to be crossed increases. 
This behaviour is shown in \Fig{sur}, where the survival probability
$P(\nu_e\rightarrow\nu_e)$ is plotted as a function of the neutrino
energy for $\MM_e=2.1\cdot 10^{-3}\eV$ and three values of $1/R$ and
$\mbb_e$. The three values of $1/R$, $0.02\eV$, $0.005\eV$ and
$0.0005\eV$, represent a sample of the different possibilities offered
by the model. The values of the brane-bulk couplings have been
determined in each case by fitting the total rates measured in the
SuperKamiokande~\cite{suzuki:99a,Suzuki2000},
Homestake~\cite{Cleveland:98a} and
Gallium~\cite{Abdurashitov:99a,Hampel:98a,Altmann:00a}
experiments. They are $\mbb_e=0.58\cdot 10^{-4}\eV$, $\mbb_e=0.35\cdot
10^{-4}\eV$, $\mbb_e=0.12\cdot 10^{-4}\eV$ respectively. The
corresponding predictions for the recoil energy spectrum in
SuperKamiokande are shown in \Fig{spe} together with the latest
data~\cite{Sobel2000}.

The dotted line corresponds to the regime $\MM_e R\ll 1$. The
probability does not depend on the precise value of $1/R$ for the
energy range shown in \Fig{sur} as long as $1/R\gtrsim 7\MM_e$. In
this case, in fact, $2 E_\nu V_e^{\text{core}} < \MM_e^2+1/R^2$, so
that the resonant mixing between the electron neutrino and the sterile
tower is significant only for the lowest state. The mixing angle with
the lowest state $\nuh_0$ is $\theta_0\simeq \mbb_e/\MM_e$. The value
of $\mbb_e$ corresponds to $\sin^2 2\theta_0\simeq 3\cdot 10^{-3}$ and
to a light neutrino mass of $1.7\cdot 10^{-6}\eV$. The phenomenology
of this case is the same as the one for a model with a single sterile
neutrino --- the Kaluza Klein origin makes no difference.
The SuperKamiokande collaboration has claimed that their data on the
day-night asymmetry and especially on the electron recoil energy
spectrum disfavour such a scenario involving oscillations into a
single sterile neutrino. This is confirmed by \Fig{spe}, where the
dotted line has the highest slope and therefore gives the worse fit of
the data\footnote{With free spectrum normalization, such a fit is
unacceptable at 95\%. A better agreement can however be found for
mixing angles at the lower border of the range allowed at 90\% CL by
the total rate fit~\cite{LRRR}.}, which is well compatible with a
flat spectrum. 

The presence of additional sterile states reduces the slope of the
predicted energy spectrum. This is illustrated by the dashed line in
\Fig{sur} and the corresponding dashed histogram in \Fig{spe}.
They both correspond to the
small $1/R$ case $1/R=0.5\cdot 10^{-3}\eV$ in which $\MM_e R >
1$. This situation is quite new because a relatively large number of
levels undergo resonant conversion. The single resonances are not
visible in the figure due to the average over the neutrino production
point in the sun. The mass of the light neutrino is now $0.9\cdot
10^{-6}\eV$. This case leads to a significant improvement with respect to
the single sterile neutrino case with the fit of the total rates (the
energy spectrum) at 65\% CL (50\% CL). Finally, the solid lines in
\Fig{sur}, \Fig{spe}
represent the intermediate case $1/R=0.005\eV$ in which, besides the
light state $\hat\nu$, three bulk states, $\nuh_0$, $\nuh_1$ and
$\nuh_2$, participate in solar oscillations. \Fig{spe} shows that
the values of the bulk and brane-bulk masses giving the best fit of
total rates (within a 25\% CL) also beautifully fit the energy
spectrum (10\% CL). Besides the three examples discussed, the whole
range of possible values of $1/R$ also includes the situation in which
$\MM_e R\sim 1$~\cite{LRRR}, that approximately reproduces the case
considered by Dvali and Smirnov~\cite{Dvali:99a}.

Unlike the energy spectrum, the day-night
asymmetry~\cite{Fukuda:98b,Suzuki2000} does not discriminate between
the different possibilities discussed. In fact, the day-night
asymmetry turns out to be small compared to the experimental errors in
all cases considered above and therefore is in reasonably good
agreement with the present data, given that SK finds only a
$1.2\,\sigma$ deviation from zero asymmetry.

For our three sample cases the values of $m_e^2R/\eV$ are of the order
of $10^{-7}$ and, hence, compatible with the supernova bounds derived
in Section~\ref{sec:SN}.

The scenario discussed leads to various experimental signatures.
First of all, a crucial test will be provided by the neutral/charged
current ratio measurement in SNO~\cite{Boger:99a}. Moreover, different
sterile neutrino scenarios can be distinguished by the shape of the
energy spectrum. As discussed, the present data already favours a
flatter spectrum, which can be obtained if more than one sterile
neutrino takes part in the oscillations. The charged current spectrum
that will be measured by SNO could provide additional
information. Finally, the measurement of a day-night asymmetry
significantly lower than zero would disfavour the scenario with many
sterile neutrinos.

\subsection{Atmospheric neutrinos}

\label{atm}
\noindent

Let us now discuss atmospheric neutrinos. As already pointed out, in a
model with small (perturbative) brane-bulk mixing, the small bulk
components of the SM neutrinos do not contribute significantly to
atmospheric neutrino oscillations. Hence, these oscillations must be
due to the three light mass eigenstates. We can, therefore, focus on
the first part of eq.~\eqref{flavour} given by
\begin{equation}
\label{3nu}
\nu_i = \sum_{k=1,2,3}U_{ik}\hat\nu_k \; .
\end{equation}
The light masses and the MNS mixing matrix $U$ are determined by the
light mass matrix $\ml$ in eq.~(\ref{mlight}). Before constructing an
explicit model, let us first discuss some phenomenological
requirements on this mass matrix suggested by atmospheric neutrino
oscillations. The simplest texture leading to maximal
$\nu_\mu\leftrightarrow\nu_\tau$ mixing is 
\begin{equation}
\label{texture}
\ml = 
\begin{pmatrix}
a & \epsilon_{e\mu} & \epsilon_{e\tau} \\
\epsilon_{e\mu} & \epsilon_\mu & 1 \\
\epsilon_{e\tau} & 1 & \epsilon_\tau
\end{pmatrix} \; m_{\mu\tau} \; ,
\end{equation}
where the $\epsilon$ parameters are much smaller than one. The
parameter $a$ is not necessarily small but it will be in our explicit
model described below. 

In the limit in which the $\epsilon$ parameters are vanishing, the muon
and tau neutrinos are superpositions of two degenerate states
$\hat\nu_2$ and $\hat\nu_3$ with mass $|m_{\mu\tau}|$. The small $\epsilon$
parameters (assumed to be real for simplicity) are necessary to generate a
mass splitting 
\begin{equation}
\Delta m_{23} \simeq \epsilon\, |m_{\mu\tau}|, \qquad \epsilon =
\epsilon_\mu+\epsilon_\tau
\end{equation}
between the degenerate states and therefore an ``atmospheric'' squared
mass difference $\dm{ATM} = \dm{23}\simeq 2\epsilon
|m_{\mu\tau}|^2$. The corresponding mixing angle $\theta_{\mu\tau}$ is
almost maximal and it is given by
$\sin^2 2\theta_{\mu\tau}\simeq 1 -(\epsilon_\mu
-\epsilon_\tau)^2/4$. For $\dm{12}\simeq\dm{13}\gtrsim\dm{ATM}$
the parameters $\epsilon_{e\mu}$, $\epsilon_{e\tau}$ are constrained
to be small by the $\nu_e$ disappearance experiments.

Notice that the simple mechanism described to generate a maximal
$\nu_\mu\leftrightarrow\nu_\tau$ mixing does not work in a three
neutrino scenario aiming at explaining at the same time the solar
neutrino data. This is because the other two squared mass differences
available turn out to be larger than the atmospheric one, unless the
parameter $a$ in~(\ref{texture}) is close to 1 and all neutrino masses
are degenerate~\cite{Barbieri:99b}. If this is not the case and all
$\epsilon<1/2$, we have, in fact,
\begin{equation}
\dm{21}\simeq\dm{31}\gtrsim|m_{\mu\tau}|^2 >
2\epsilon |m_{\mu\tau}|^2 = \dm{ATM} \; . 
\end{equation}
As a consequence, there is no room for the small $\dm{}$ required by
the solar data. In our case, of course, such a small $\dm{}$ is not
needed since the solar neutrino problem is solved by oscillations into
bulk neutrinos.

If the texture~(\ref{texture}) accounts for atmospheric neutrino
oscillations, the two degenerate neutrinos can provide a
cosmologically significant source of dark matter. Moreover, the
electron neutrino can oscillate into muon or tau neutrinos with a
squared mass difference $\dm{} > \dm{ATM}$ and small amplitudes
$4\epsilon^2_{e\mu}$, $4\epsilon^2_{e\tau}$ respectively. If the
parameter $a$ is small, we have $\D m\simeq m_{\m\t}$. It is obviously
tempting to associate such oscillations with the signal observed by
LSND. This then determines the size of $m_{\m\t}$ to be
$|m_{\mu\tau}|\simeq\sqrt{\dm{LSND}}$. The MiniBooNE experiment will
test this possibility and will cover a relevant portion of the
parameter space for such short-baseline oscillations. A short-baseline
neutrino factory could further extend the sensitivity in the
$\epsilon_{e\mu}$ mixing parameter~\cite{BGRW}.

\interskip

Let us now see how the texture~(\ref{texture}) can be obtained in a
model in which the single family discussion of solar neutrino
oscillations given above can be embedded. We work in the
context of the $\ZZ$ invariant models with Majorana bulk mass
described in Subsection~\ref{sec:mm}. 

A simple way to obtain the zeroth order form of \eq{texture} in which
$a$ and the $\epsilon$ parameters vanish is to assume three
bulk fermions $\Psi_e$, $\Psi_\mu$, $\Psi_\tau$, a diagonal
brane-bulk coupling $m$ and a  Majorana mass matrix $M$ with
two off-diagonal non vanishing entries. Concretely,
\begin{equation}
\label{nutau}
\begin{matrix}
\phantom{
 {\mbb} = \,
\begin{matrix}
\nu_\tau
\end{matrix}\,\;
} 
\begin{matrix}
\Psi_e & \Psi_\mu & \Psi_\tau 
\end{matrix} \\[1mm]
{\mbb}= \,
\begin{matrix}
\nu_e \\
\nu_\mu \\ 
\nu_\tau \\ 
\end{matrix}
\,
\begin{pmatrix}
  \phantombox{$\Psi_e$}{$\mbb_e$} &\phantombox{$\Psi_\mu$}{$0$}
  & \phantombox{$\Psi_\tau$}{$0$} \\ 
  0 & \mbb_\mu & 0 \\
  0 & 0 & \mbb_\tau
\end{pmatrix}
\end{matrix}
\qquad
\begin{matrix}
\phantom{
 \MM = \,
\begin{matrix}
\Psi_\tau
\end{matrix}\,\;
} 
\begin{matrix}
\Psi_e & \phantombox{$\MM_{\mu\tau}$}{$\Psi_\mu$} &
\phantombox{$\MM_{\mu\mu}$}{$\Psi_\tau$} 
\end{matrix} \\[1mm]
\MM = \,
\begin{matrix}
\Psi_e \\
\Psi_\mu \\ 
\Psi_\tau \\ 
\end{matrix}
\,
\begin{pmatrix}
\phantombox{$\Psi_e$}{$0$} & 0 & 0 \\
0 & 0 & \MM_{\mu\tau} \\
0 & \MM_{\mu\tau} & 0
\end{pmatrix}& \hspace*{-1ex}{,}
\end{matrix}
\end{equation}
where $\MM_{\mu\tau}$ can be made real by a phase-redefinition of the
bulk fields $\Psi_\mu$, $\Psi_\tau$. We note that, for a direct
application of the results of Subsection~\ref{sec:mm} to this case, we
should first diagonalize the matrix $M$ and rewrite the brane-bulk
coupling $m$ in the corresponding basis. Alternatively, one can use
\eq{seesaw} with $\mbb$ and $\MM$ as in \eq{nutau}.  Then, we find
from eq.~\eqref{mlight} that the light mass matrix $\ml$ has indeed
the structure~\eqref{texture} with vanishing parameters $a$, $\e$ and
the overall scale $m_{\m\t}$ given by
\begin{equation}
\label{mnutau}
m_{\mu\tau} \simeq -\pi R \mbb_\mu\mbb_\tau\coth (\pi R \MM_{\mu\tau}) \; .
\end{equation}
The bulk components of $\nu_\mu$ and $\nu_\tau$ are small if 
\begin{equation}
\label{con1}
\pi R |\mbb_\mu|^2\coth (\pi R \MM_{\mu\tau}) \ll  \MM_{\mu\tau}
\quad\text{and}\quad 
\pi R |\mbb_\tau|^2\coth (\pi R \MM_{\mu\tau}) \ll  \MM_{\mu\tau} \; ,
\end{equation}
which implies that
\begin{equation}
\label{con2}
|m_{\mu\tau}| \ll \MM_{\mu\tau} \; .
\end{equation}

The form~(\ref{nutau}) of the brane-bulk mixing and of the Majorana
mass matrix can be enforced by means of symmetries acting both on
brane and bulk fields. For example, let us introduce three individual
lepton numbers $L_{e,\mu,\tau}$ for each generation with the bulk
fermions $\Psi_{e,\mu,\tau}$ carrying the same charge as the corresponding
neutrinos on the brane. Let us further require that $L_\mu - L_\tau$
and $L_e$ are both conserved. This leads exactly to the structure
given in eq.~\eqref{nutau}. In order to generate the small $\e$
corrections to the texture~\eqref{texture} one has to break these
symmetries. This can be done by small off-diagonal brane-bulk
couplings or small Majorana masses replacing the zero entries in the
matrices~\eqref{nutau}. 

To see this in more detail, let us first discuss how solar neutrino
oscillations along the lines of Section~\ref{sec:solar} can be
embedded into the model. As a consequence of \eqs{nutau}, the electron
neutrino only mixes with the bulk field $\Psi_e$. The solar neutrino
phenomenology described in Section~\ref{sec:solar} follows once $L_e$
is broken by a non-vanishing Majorana mass $\MM_e$. The smallness of
$\MM_e \sim(2\dash 3)\cdot 10^{-3}\eV$ compared to
$\MM_{\mu\tau}\gg|m_{\mu\tau}| > \sqrt{\dm{ATM}}$ is then a
consequence of the approximate conservation of $L_e$. Analogously, a
small breaking of $L_\mu-L_\tau$ can produce the small diagonal
entries $\epsilon_\mu$ and $\epsilon_\tau$ necessary to generate the
atmospheric $\dm{}$. This can be accomplished by switching on the
$L_\mu-L_\tau$ breaking and $L_e$ conserving entries in the bulk
Majorana matrix or in the brane-bulk mixing matrix. Once $L_e$ and
$L_\mu-L_\tau$ are both broken, the $\nu_e$-$\nu_\mu$ mixing
in~(\ref{texture}) can be easily generated in the same way. In
general, $L_\mu-L_\tau$ can be broken both by bulk Majorana mass terms
and by brane-bulk couplings. The most general case, corresponding to
replacing all zero entries in the texture~\eqref{nutau} by small
corrections, leads to a quite complicated mass matrix for the light
neutrinos. As an illustration, let us consider a somewhat simplified
case where we break $L_\mu-L_\tau$ by brane-bulk terms only. That is,
instead of eq.~\eqref{nutau} we now consider matrices
\begin{equation}
 \label{nutau1}
 m = \left(\begin{array}{ccc}m_e&\d m_{e\m}&\d m_{e\t}\\
     \d m_{\m e}&m_\m&\d m_{\m\t}\\
     \d m_{\t e}&\d m_{\t\m}&m_\t
     \end{array}\right)\qquad
 M = \left(\begin{array}{ccc}M_e&0&0\\
     0&0&M_{\m\t}\\
     0&M_{\m\t}&0
     \end{array}\right)
\end{equation}
where $\d m_{e\m}$, $\d m_{e\t}$ and $\d m_{\m\t}$ are small
corrections. Expanding to first order in these corrections and using
the parameterization~\eqref{texture}, we find for the $\e$
perturbations to the mass matrix
\bea
 \e_\m &=& \frac{2\d m_{\t\m}}{m_\t} \\
 \e_\t &=& \frac{2\d m_{\m\t}}{m_\m} \\
 \e_{e\m} &=& \frac{m_e\coth (\p RM_e)\d m_{e\m}+m_\m\coth
 (\p RM_{\m\t})\d m_{\t e}}{m_\m m_\t\coth (\p R M_{\m\t})} \\
 \e_{e\t} &=& \frac{m_e\coth (\p RM_e)\d m_{e\t}+m_\t\coth
 (\p RM_{\m\t})\d m_{\m e}}{m_\m m_\t\coth (\p R M_{\m\t})}\; .
\eea
The atmospheric square mass difference is then related to the
corrections by
\begin{equation}
 \e = \frac{\D m_{\rm ATM}^2}{2|m_{\m\t}|^2} =
 2\left(\frac{\d m_{\m\t}}{m_\m}+\frac{\d m_{\t\m}}{m_\t}\right)\; .
\end{equation}
The amount of $L_\mu-L_\tau$ breaking needed for a correct account of
atmospheric neutrino oscillations is, therefore, variable since it
depends on $m_{\m\t}$ which is not fixed by solar or atmospheric
data. If, in addition, we require an explanation of the LSND signal we
have $m_{\m\t}\simeq\sqrt{\dm{LSND}}$ resulting in a $L_\mu-L_\tau$
breaking $\e$ of the order $\sqrt{\dm{ATM}/\dm{LSND}}$. For a
hierarchical structure of the brane-bulk mass matrix, that is
$\mbb_e\ll\mbb_\m\ll\mbb_\t$ and $\d m_{\t e}\lesssim\d m_{\m e}$,
the $e$--$\m$ mixing angle $\theta_{e\m}$ is given by
\begin{equation}
 \theta_{e\m} = \frac{\e_{e\m}+\e_{e\t}}{\sqrt{2}}\simeq
 \frac{\d m_{\m e}}{\sqrt{2}m_\m}\; . 
\end{equation}
This angle should then be in the LSND range which can be easily
accommodated by choosing $\d m_{e\m}$. Notice that such a
small $\nu_e$-$\nu_\mu$ mixing can also generate a small mixing
between the electron neutrino and the towers of sterile
neutrinos associated with $\Psi_{\mu,\tau}$, heavier than the one we
have used to deplete the solar neutrino flux. Due to the large
mass splitting between the lightest electron neutrino and the lowest mode
of the heavy towers, such a small mixing can not be MSW-enhanced and
consequently does not affect the solar neutrino
phenomenology. However, it may give rise to a non-negligible
contribution to the light electron neutrino mass. 

It is clear, that a parallel discussion can be carried out for
$L_\mu-L_\tau$ breaking induced by bulk Majorana mass terms.
However, the above example is sufficient to show that all oscillation
data can be explained in terms of our model. We, therefore, will not
carry this out explicitly.

We have already mentioned that parameters in the model
associated with the first generation are compatible with the supernova
bound. However, we also have to worry about the supernova bound due to resonant
conversion of $\n_\m$ and $\n_\t$ into their respective Kaluza-Klein
states. As the above equations show, the particle
phenomenology of the model is not very sensitive to $M_{\m\t}$
as long as this mass is larger than $m_{\m\t}$ and smaller than
the fundamental scale. Therefore, we can choose $M_{\m\t}$ larger than
$\sqrt{2EV}$ where $V$ is the matter potential in the core of the
supernova and $E$ is the neutrino energy. This avoids any problem with
the supernova bound due to conversion of $\n_\m$ and $\n_\t$ into
their associated bulk states. As discussed, the large value of
$M_{\m\t}$ (as compared to $M_e$) can be attributed to the approximate
conservation of $L_e$.

\medskip

To summarize, in the framework of a $\ZZ$-symmetric theory we have
accommodated solar neutrino oscillations, naturally maximal
$\nu_\mu\leftrightarrow\nu_\tau$ oscillations, two degenerate
neutrinos providing a cosmologically significant source of dark matter
and $\nu_e\leftrightarrow\nu_\mu$ oscillations with a $\dm{}$ larger
than the atmospheric one, that can be used to accommodate the LSND
signal. The spectrum of parameters in the bulk Majorana mass matrix
and the brane-bulk mixing matrix~\eqref{nutau1} we need is the following.
\begin{itemize}
\item
The value of $1/R$ should be compatible with its lower experimental
limit emerging from Cavendish-type experiments.
\item
To get the best agreement with the SuperKamiokande energy spectrum,
$1/R$ should be less that $10^{-2}\eV$, in such a way that more than a
sterile state is involved. 
\item
To explain the solar neutrino deficit $L_e$ has to be broken by
an electron bulk Majorana mass
$\MM_e \sim (2\dash 3)\cdot 10^{-3}\eV$. Furthermore, we need
the electron brane-bulk coupling to be of the order
$\mbb_e = \ord{10^{-(4\dash 5)}\eV}$.
\item
If the LSND signal has to be explained the parameter $|m_{\mu\tau}|$
as given in eq.~\eqref{mnutau} should have the value $|m_{\mu\tau}| =
\sqrt{\dm{LSND}}=\ord{1\eV}$. This requires that the quantity
$\sqrt{\mbb_\mu\mbb_\tau}$ is close to the geometric average of
$\sqrt{\dm{LSND}}$ and the smallest between $\MM_{\mu\tau}$ and
$R^{-1}$.
\item
We need $\MM_{\mu\tau} \gg |m_{\m\t}|$, in order to have a small bulk
component in $\nu_\mu$ and $\nu_\tau$ and a quasi-unitary light mixing
matrix $U$. A value of $M_{\m\t}$ larger than $\sqrt{2EV}$ avoids any
potential problems with the conversion of $\n_\m$ and $\n_\t$ into
their associated bulk states in the supernova. Also the hierarchy
between $\mbb_\mu$ and $\mbb_\tau$ must be smaller than the hierarchy
between $|m_{\m\t}|$ and $\MM_{\mu\tau}$.
\item 
To explain the atmospheric squared mass difference we need
to break $L_\mu-L_\tau$. If this is done be brane-bulk mixing
and we use the LSND result the level of this breaking is
specified by the condition
$\d m_{\m\t}/m_\m\simeq \D m_{\rm ATM}^2/\D m_{\rm LSND}^2$.
\item
In order to get the $e$--$\m$ mixing required by LSND we need to
consider terms breaking $L_e$ and $L_\mu-L_\tau$. If this is achieved
by perturbing the brane-bulk mixing the level of this breaking is
determined by $\theta_{e\m} = \d m_{\m e}/m_\m$. A smaller value of
$\theta_{e\mu}$ than needed to account for the LSND signal could be
within the reach of MiniBooNE or a short-baseline neutrino factory.

\end{itemize}

Finally, let us mention that there are alternative models within the
general class considered in this paper which provide similar
possibilities to explain the various oscillation phenomena. For
example, in the context of the models with $U(1)$ lepton number
symmetry and Dirac bulk masses which we have briefly discussed,
an analogous (and complementary) scheme using the brane-bulk mixings
$\mbbc$ as a source of $U(1)$-breaking and light neutrino masses can
be constructed~\cite{LRRR}. Then, the overall $U(1)$ can be considered as a
subgroup of the individual lepton number groups and the model
with $L_\mu - L_\tau$ and $L_e$ symmetry arises at an intermediate
stage in the necessary breaking of these individual lepton numbers.

\section{Conclusion}

In this paper, we have systematically studied five-dimensional
brane-world models of neutrino physics based on flat
compactifications. Motivated by the expectation from string theory, we
have particularly focused on the effect of bulk mass terms for the 
sterile neutrinos. For such models we have presented a number of
general results. We have pointed out that there are four different
types of bulk mass terms, namely scalar Dirac and Majorana masses as
well as vector-like Dirac and Majorana masses. While the scalar masses
can be thought of as being generated by the VEV of a bulk scalar field
and respect five-dimensional Lorentz invariance the vector masses are due
to the VEV of the fifth component of a bulk vector field and break
five-dimensional Lorentz invariance. We have analyzed the
(unperturbed) spectrum of bulk Kaluza-Klein states in the presence of
these mass terms and have found a crucial difference between scalar
and vector-like masses. Whereas scalar masses constitute a lower bound on
the spectrum this is not the case for vector-like masses. For
arbitrary values of these masses (and in the absence of scalar masses)
there will always be states with masses of order $1/R$. In the
presence of small brane-bulk couplings, we have presented perturbative
results based on a see-saw type approximation. In particular, we have
shown in general that, in this approximation, the electroweak
eigenstates are mainly given by a superposition of three light mass
eigenstates with small admixtures of massive Kaluza-Klein modes. We
have also shown that the relevant parameters associated with these two
sectors can be controlled quite independently by choosing the various
mass terms in the action appropriately. This opens up the possibility
of associating some of the observed oscillation phenomena with
``standard oscillations'' into the light states and some others with
oscillations into heavy bulk states. To complement the perturbative
approach, we have also presented a general formalism and results
for the exact diagonalization of the mass matrix. 

These general results have been applied to construct a specific
brane-world model compatible with all presently observed oscillation
phenomena. The model is based on a $\ZZ$ orbifolding and broken
$L_e$ and $L_\m - L_\t$ symmetries. The solar neutrino deficit is
explained by small mixing angle oscillations into the tower of heavy
bulk states. It has been shown that this, contrary to the case of
small mixing angle oscillations into a single sterile neutrino, is
compatible with recent SuperKamiokande results. The model also
naturally leads to a maximal $\nu_\m\leftrightarrow\nu_\tau$ mixing
that accounts for the atmospheric neutrino result. In addition, it
allows to accommodate the LSND observation and may lead to a significant
neutrino dark matter component.

We have also analyzed constraints from supernova energy loss on
such brane-world models and have demonstrated that our specific model
is consistent with these constraints.

In summary, we have shown that the inclusion of bulk mass terms leads
to a rich structure of brane-world models for neutrino masses which
can be used to construct models compatible with all known experimental
results.

\noindent

\section*{Acknowledgements}

We would like to thank Subir Sarkar for useful discussions.
This work is supported by the TMR Network under the EEC Contract
No.~ERBFMRX--CT960090. One of us (PR) was Dr.~Lee Visiting Fellow
at Christ Church College, Oxford, and is partially supported by
DOE under grant number DE-FG05-86-ER40272.

%%%%%%%%%%%%%%%%%%%%%%%%%%%%%%%%%%%%%%%%%%%%%%%%%%%%%%%%%%%%%%%%%%%%%%%

\appendix

\section{Spinor conventions in four and five dimensions}

\label{app:spinor}

\noindent

In this Appendix, we collect the properties of five-dimensional gammma
matrices and spinors which we are using in the main part of the paper.

The five-dimensional gamma matrices $\g_\a$, where
$\a ,\b ,\cdots = 0,1,2,3,4$, satisfy the Clifford algebra
\begin{equation}
 \{\g_\a ,\g_\b\} = 2\eta_{\a\b}=2\,\text{diag}(1,-1,-1,-1,-1)\; .
\end{equation}
Note, that we are using the ``mostly minus'' Minkowski metric
throughout. As usual, the generators $\Sigma_{\a\b}$ of the Lorentz group
$SO(4,1)$ can be obtained by
\begin{equation}
 \Sigma_{\a\b} = \frac{i}{4}[\g_\a ,\g_\b ]\; .
\end{equation}
The hermitian conjugate of the gamma matrices is generated by $\g_0$
according to
\begin{equation}
 (\g_\a )^\dagger = \g_0\g_\a\g_0\; .
\end{equation}
Similarly, to express the complex conjugate and the transpose of the
gamma matrices we introduce a matrix $B$ and a charge conjugation
matrix $C$. Their defining properties are
\bea
 (\g_\a )^* &=& B\g_\a B^{-1} \\
 (\g_\a )^T &=& C\g_\a C^{-1}\; .
\eea
Obviously, this implies that $C=B^T\g_0$. We note, that the matrices
$B$ and $C$ square to $-1$ unlike their four-dimensional counterparts
which square to one. 

A five-dimensional Dirac spinor $\Psi$ has four complex components.
Under infinitesimal Lorentz transformations it transforms as
\begin{equation}
 \d\Psi = \frac{i}{2}\o_{\a\b}\Sigma_{\a\b}\Psi\; .
\end{equation}
We define the conjugated spinor $\bar{\Psi}$ and the charge conjugated
spinor $\Psi^c$ by
\begin{equation}
 \bar{\Psi} = \Psi^\dagger\g_0\; ,\qquad
 \Psi^c = B^{-1}\Psi^* = C\bar{\Psi}^T\; .
\end{equation}
Note that, under infinitesimal Lorentz transformations, $\Psi^c$
transforms in the same way as $\Psi$. Furthermore, we note that
$(\Psi^c)^c=-\Psi$. The minus sign in this relation is due to the
above mentioned fact that $B^2=C^2=-1$ and precludes the existence of
five-dimensional Majorana spinors.

For our purpose of dimensional reduction to four dimensions, we need
to make contact between the above five-dimensional spinors and Weyl
spinors. To do this, it is useful to introduce explicit
representations for the gamma matrices. We choose
\begin{equation}
 \g_\m = 
\begin{pmatrix}
0 & \s_\m \\
\bar{\s}_\m & 0
\end{pmatrix} \; ,
 \qquad
 \g_4 = \g_0\g_1\g_2\g_3 = 
\begin{pmatrix}
-i\um_2 & 0 \\
0 & i\um_2
\end{pmatrix}\;
\end{equation}
where $\s_\m =(\um_2,-\s_i)$, $\bar{\s}_\m =(\um_2,\s_i)$ and $\s_i$
are the Pauli matrices. Furthermore, we have the matrix
$\g_5$ defined as usual by
\begin{equation}
 \g_5 = i\g_4=
\begin{pmatrix}
\um_2 & 0 \\
0 & -\um_2
\end{pmatrix}
\end{equation}
and satisfying $(\g_5)^2=1$. For the matrices $B$ and $C$ we can
choose, in this representation
\begin{equation}
 B = \g_2\g_4 = -\begin{pmatrix}0&\e\\\e&0\end{pmatrix}\;
 ,\qquad
 C = \g_2\g_0\g_4 =
 \begin{pmatrix}\e&0\\0&\e\end{pmatrix}
\end{equation}
where $\e =i\s_2$ is the two-dimensional epsilon symbol.

A five-dimensional Dirac spinor $\Psi$ is decomposed into two
left-handed Weyl spinors $\x$ and $\eta$ as
\begin{equation}
 \Psi = \begin{pmatrix}\bar{\x}\\\eta\end{pmatrix}\; .
\end{equation}
The conjugated spinor $\x$ is defined by $\bar{\x} = \e\x^*$.
Using the above explicit representation, it is easy to obtain the
relations
\begin{equation}
 \bar{\Psi} =
 \begin{pmatrix}\eta^\dagger&-\x^T\e\end{pmatrix}\;
 ,\qquad
 \Psi^c = \begin{pmatrix}\bar{\eta}\\-\x\end{pmatrix}\;
 ,\qquad
 \bar{\Psi}^c =
 \begin{pmatrix}-\x^\dagger&-\eta^T\e\end{pmatrix}
\end{equation}
which will be useful for the dimensional reduction to four dimensions.
To write our four-dimensional effective action in a compact form we
furthermore adopt the following notation for Weyl mass terms
\begin{equation}
 \x_1\x_2 \equiv \x_1^T\e\x_2\; .
\end{equation}
where $\x_1$ and $\x_2$ are two (left-handed) Weyl spinors.

\section{Exact diagonalization}

\label{app:diag}
\noindent
We first consider the general case of three standard model neutrinos
$\bnu=(\nu_e,\nu_\mu,\nu_\tau)^T$ and $\kappa$ ``bulk'' neutrinos
$\bchi=(\chi_1\ldots\chi_\kappa)^T$. The associated Majorana mass
matrix is of the general form
\begin{equation}
\begin{matrix}
\phantom{
 {\cal M} = \,
\begin{matrix}
\eta_n 
\end{matrix}\,\;
} 
\begin{matrix}
\bnu^T & \bchi^T 
\end{matrix} \\[1mm]
{\cal M}= \,
\begin{matrix}
\bnu \\ 
\bchi \\ 
\end{matrix}
\,
\begin{pmatrix}
  \phantombox{$\bnu^T$}{$0$} & \phantombox{$\bchi^T$}{$\bv^T$} \\
  \bv & \bD 
\end{pmatrix} & \hspace*{-1ex}{,}
\end{matrix}
 \label{Mapp}
\end{equation}
where $\bv$ is a $\kappa\times 3$ matrix and $\bD$ is a
$\kappa\times\kappa$ symmetric matrix. When the set of $\bchi$ fields
is allowed do be infinite, the results can be applied to the
diagonalization of the mass matrix~\eqref{mmatrix}, originating from
the five-dimensional brane-world theory, by making the obvious
identifications
\globallabel{id}
\begin{gather}
\bnu = (\nu_e,\nu_\mu,\nu_\tau)^T \qquad \bchi =
{(\xi_n,\eta_n)_{n\in\Z}}^T \mytag \\
\bv^T = (\mbb^T,{\mbbc}^T)_{n\in\Z} \qquad \bD =
(\cM_n\delta_{-m,n})_{n,m\in\Z}.
\mytag
\end{gather}
While, in this paper, we focus on cases where $\bD$ has such a block
structure, the general formulae below also apply to more complicated
cases. A more complicated $\bD$ is generated if the bulk mass
terms effectively depend on the coordinates of the additional dimension.
Such an explicit coordinate dependence may arise in orbifold models or
due to a non-flat vacuum solution.

Since, with the usual approximations, the equations for neutrino
propagation only involve $\cM^\dagger\cM$, we actually
diagonalize the matrix
\begin{equation}
\label{Meff}
\cM^2_{\text{eff}} = \cM^\dagger\cM + 
\begin{pmatrix}
A & 0 \\
0 & 0
\end{pmatrix}
=
\begin{pmatrix}
\bv^\dagger\bv+A & \bv^\dagger \bD \\
\bar\bD\bv & \bar\bD\bD +\bar\bv\bv^T
\end{pmatrix} \; ,
\end{equation}
where the hermitian matrix $A = \sqrt{2} E_\nu V$ takes matter effects
into account. As usual, $E_\nu$ represents the neutrino energy and $V$
the matter induced potential.  The eigenvalue equations for the
generic mass eigenstate
\begin{equation}
\label{eigen}
\hat\nu = \balpha^\dagger\bnu + \bbeta^\dagger\bchi
\end{equation}
(in obvious matrix notations) are
\begin{equation}
\label{eieq}
\left\{
\begin{aligned}
\mbox{}& (\bv^\dagger\bv+A)\balpha + \bv^\dagger\bD\bbeta = \lambda^2\alpha \\
\mbox{}& \bar\bD\bv\balpha + \bar\bD\bD\bbeta +\bar\bv\bv^T\bbeta =
\lambda^2 \bbeta 
\end{aligned}\; ,
\right.
\end{equation}
where $\lambda^2$ is the corresponding real (possibly negative)
eigenvalue. 

As shown in Appendix~\ref{app:majorana} and further explained in the
main text, it often happens that some of the
eigenvectors of the matrix $\bar\bD\bD$ are also eigenvectors of the
matrix $\cM$ and as such decouple from the light neutrinos
$\bnu$. The vector space generated by such ``unperturbed'' mass
eigenstates coincides with the maximal space of vectors in the form
\eq{eigen} such that $\balpha=0$, $\bbeta$ is orthogonal to both
$\bar\bv$ and $\bar\bD\bv$ and the space is invariant under
$\bar\bD\bD$. In other words the unperturbed eigenstates are in the form
$\hat\nu_{\rm unp} = \bbeta^\dagger\bchi$, with $\bbeta$ solving the
following set of equations
\begin{equation}
\label{sysunpert}
\left\{
\begin{aligned}
\mbox{}& \bar\bD\bD\bbeta = \lambda^2\bbeta \\
\mbox{}& \bv^\dagger\bD\bbeta = 0 \\
\mbox{}& \bv^T\bbeta = 0
\end{aligned}
\right.\; .
\end{equation}
Let us now derive the eigenvalue equations for the ``perturbed''
eigenstates. In this case, it is reasonable to assume that $\lambda^2$
is not an eigenvalue of $\bar\bD\bD$, so
that $\lambda^2-\bar\bD\bD$ can be inverted~\footnote{The eigenvalue
of a perturbed eigenstate could ``accidentally'' be an eigenvalue
of $\bar{\bD}\bD$ as well. Here, we do not consider such pathological
cases explicitly.}. Then, using \eqs{eieq} $\bv^\dagger\bD\bbeta$ and
$\bbeta$ can be expressed in terms of $\tilde{\balpha}\equiv\bv^T\bbeta$
and $\balpha$. In particular,
\begin{equation}
\label{beta}
\bbeta = \frac{1}{\lambda^2-\bar\bD\bD} \left[
\bar\bD\bv\balpha + \bar\bv (\tilde{\balpha})\right] \; .
\end{equation}
By taking appropriate projections, one then
gets the following equations for $\tilde{\balpha}=\bv^T\bbeta$ and
$\balpha = (\lambda^2-\bv^\dagger\bv -A)^{-1}\bv^\dagger\bD\bbeta$:
\begin{equation}
\label{projections}
\left\{
\begin{aligned}
\mbox{}& \left(\mathbf{1} - 
\bv^T \frac{1}{\lambda^2-\bar\bD\bD} \bar\bv
\right) \tilde{\balpha} = \bv^T \frac{1}{\lambda^2-\bar\bD\bD}
\bar\bD \bv \balpha \\
\mbox{}& \left(\lambda^2 -\bv^\dagger\bv -A - \bv^\dagger\bD
\frac{1}{\lambda^2-\bar\bD\bD} \bar\bD 
\bv \right) \balpha =
\bv^\dagger\bD \frac{1}{\lambda^2-\bar\bD\bD} \bar\bv \tilde{\balpha}
\end{aligned}
\right.\; .
\end{equation}
\Eqs{projections} can have two different sets of solutions. In some cases
there can be ``decoupled solutions'' corresponding to eigenstates
decoupled from the SM eigenstates, that is $\balpha=0$. The decoupled
states of squared mass $\lambda^2$ are therefore in the form
\begin{equation}
\label{statedec}
\hat\nu_{\text{dec}} = \tilde{\balpha}^\dagger\bv^T
\frac{1}{\lambda^2-\bar\bD\bD} \bchi \; ,
\end{equation}
where $\tilde{\balpha}=\bv^T\bbeta$ must satisfy
\begin{equation}
\label{sysdec}
\left\{
\begin{aligned}
\mbox{}& \left(\mathbf{1} - 
\bv^T \frac{1}{\lambda^2-\bar\bD\bD} \bar\bv \right) \tilde{\balpha} = 0 \\
\mbox{}& 
\bv^\dagger\bD \frac{1}{\lambda^2-\bar\bD\bD} \bar\bv \tilde{\balpha} = 0
\end{aligned}
\right.\; .
\end{equation}
Although such decoupled (but not unperturbed) eigenstates are not
always present, we will see in Appendix~\ref{app:dirac} an example
in which they are relevant. Also note that the equations for both the
unperturbed and the decoupled states do not depend on matter effects
encoded in $A$. Therefore, these states are insensitive to matter
effects as one would intuitively expect.

The generic solutions of \eqs{projections} correspond to
states which couple to the SM eigenstates $\bnu$ and, therefore, have
$\balpha\neq0$. In this case, if follows from \eqs{projections} that
$\lambda^2$ is an eigenvalue if the following equation has a non-trivial
solution for $\balpha$:
\begin{multline}
\label{eigenvalues}
\left[ \lambda^2 \left(\mathbf{1} - \bv^\dagger \frac{1}{\lambda^2-\bD\bar\bD}
\bv \right) -A \right] \balpha \\ = \bv^\dagger \bD
\frac{1}{\lambda^2-\bar\bD\bD} \bar\bv \left( \mathbf{1} - 
\bv^T \frac{1}{\lambda^2-\bar\bD\bD} \bar\bv \right)^{-1} \bv^T
\frac{1}{\lambda^2-\bar\bD\bD} \bar\bD\bv \balpha \; .
\end{multline} 
In the case of a single family and no matter effects this
eigenvalue equation becomes
\begin{equation}
\label{eigenvalues1}
\lambda^2 \left| 1 - \bv^T \frac{1}{\lambda^2-\bar\bD\bD} \bar\bv \right|^2 =
\left| \bv^\dagger \bD
\frac{1}{\lambda^2-\bar\bD\bD} \bar\bv \right|^2 \; .
\end{equation} 

The eigenspace corresponding to the eigenvalue $\lambda^2$ is parameterized
by the values of $\balpha$ solving the above equation. By using
\eq{beta} with $\tilde{\balpha}$ obtained from the first of 
\eqs{projections} we find the following mass eigenstate
associated with $\lambda^2$ and $\balpha$.
\begin{equation}
\label{eigenvectors}
\hat\nu = \balpha^\dagger
\left[ \nu + \bv^\dagger\bD\left[
\mathbf{1} +\frac{1}{\lambda^2-\bar\bD\bD} \bar\bv \left(
\mathbf{1} - \bv^T \frac{1}{\lambda^2-\bar\bD\bD} \bar\bv \right)^{-1}
\bv^T \right] \frac{1}{\lambda^2-\bar\bD\bD} \chi \right] \; .
\end{equation} 

We now specialize these results to the case of a mass matrix $\bD$
with block structure as in \eqs{id}, a single SM family $\nu$ and
a single bulk fermion $(\xi_n,\eta_n)_{n\in\Z}$. This is done
for two cases, namely for a Dirac and a Majorana bulk mass term.

\subsection{Dirac bulk masses}

\label{app:dirac}
\noindent
We first consider the simple case with vanishing Majorana-type terms,
that is $\MM_S = \MM_V = 0$ and $\mbbc = 0$, and a Dirac bulk mass term with a
scalar components only, that is $\MD\equiv\MD_S$, $\MD_V = 0$. Such a
pattern is realized in the context of a model with $U(1)$ lepton
number symmetry. Furthermore, we also include matter effects.

The unperturbed  eigenstates can be derived from \eqs{sysunpert}. For each
$n\geq 1$ there are two unperturbed eigenstates with mass
$\sqrt{\MD^2+n^2/R^2}$, namely
\begin{equation}
\label{diracunpert}
\frac{\xi_{n}-\xi_{-n}}{\sqrt{2}} \qquad\text{and}\qquad
\frac{e^{-i\phi_n}\eta_{n}-e^{i\phi_n}\eta_{-n}}{\sqrt{2}} \; .
\end{equation}
Here, the phase $\phi_{n}$ is defined by $\MD+in/R = e^{i\phi_{n}}
\sqrt{\MD^2+n^2/R^2}$. 

In this model decoupled eigenstates are present as well. In fact, as a
consequence of the U(1) symmetry, the second equation
in~(\ref{sysdec}) is automatically satisfied. This is a very
particular property of the $U(1)$ symmetric model and will generically
not be the case for other model. The first equation in~(\ref{sysdec})
then leads to the eigenvalue equation
\begin{equation}
\label{diracvaldec}
\pi R |\mbb|^2 \cot\pi R \sqrt{\lambda^2-\MD^2} =
\sqrt{\lambda^2-\MD^2} \; .
\end{equation}
For each $\l$ solving this equation one obtains a decoupled state of
the form
\begin{equation}
\label{diracstatedec}
\hat\nu_{\text{dec}} = \frac{1}{\sqrt{N}} \sum_{n\in\Z}
\frac{\mbb\lambda}{\lambda^2-\MD^2-n^2/R^2}\, \xi_n \; ,
\end{equation}
The normalization factor $N$ in \eq{diracstatedec} is given by
\begin{equation}
\label{nor1}
2N = \frac{\lambda^2}{|\mbb|^2} +
\frac{\lambda^2}{\lambda^2-\MD^2}\left(1+\pi^2 R^2 |\mbb|^2\right) \; .
\end{equation}
The equation for the masses $\lambda$ of the coupled
eigenstates is given by
\begin{equation}
\label{diracval}
\lambda^2 \frac{\pi R |\mbb|^2}{\sqrt{\l^2-\MD^2}} \cot \pi R
\sqrt{\l^2-\MD^2} +A = \lambda^2 \; .
\end{equation}
This eigenvalue equation has a single solution with $\lambda^2 <
\MD^2$, as can be seen by using the relation $\cot(\sqrt{-x})/\sqrt{-x} = -
\coth(\sqrt{x})/\sqrt{x}$. In vacuum, we have
$\lambda^2 = 0$, as a consequence of the U(1) symmetry.
The associated state with mass $\lambda^2$ reads
\begin{equation}
\label{diracstate}
\hat\nu = \frac{1}{\sqrt{N}} \left[
\nu +
\sum_{n\in\Z}\frac{\bar\mbb(\MD-in/R)}{\lambda^2-\MD^2-n^2/R^2} \,
\eta_n \right ]\; .
\end{equation}
The normalization factor $N$ in \eq{diracstate} is given
by~\footnote{In vacuum the normalization of the light state is
$N=1+\p R|m|^2\coth (\p R\m )/\m$.}
\begin{equation}
\label{nor2}
2N = \frac{(\lambda^2 - A)^2 + 2 A |\mbb|^2}{\lambda^2|\mbb|^2} +
\frac{\lambda^2 (1+\pi^2 R^2 |\mbb|^2) - A}{\lambda^2-\MD^2} \; .
\end{equation}
In vacuum, the masses of the coupled eigenvalues coincide with the
masses of the decoupled ones, with the exception of $\lambda=0$ which
is only a coupled eigenvalue. Also, the normalization factors in
eqs.~(\ref{diracstatedec}) and~(\ref{diracstate}) coincide in vacuum.
Coupled and non-coupled states can be combined to form left and right-handed
components of Dirac fermions.

\subsection{Majorana bulk masses}

\label{app:majorana}
\noindent
We now consider the case of a vanishing Dirac mass, $\MD = 0$ and
furthermore assume that $\mbbc = 0$ and $\MM_V=0$. We keep the scalar
Majorana mass $\MM\equiv\MM_S$. This theory is
invariant under the $\ZZ$ symmetry.

The unperturbed  eigenstates can be derived from \eqs{sysunpert}. For each
$n\geq 1$ there are two unperturbed eigenstates with mass
$\sqrt{\MM^2+n^2/R^2}$ corresponding to the $\ZZ$-odd combinations
\begin{equation}
\label{majounpert}
\frac{\xi_{n}-\xi_{-n}}{\sqrt{2}} \qquad\text{and}\qquad
\frac{\eta_{n}+\eta_{-n}}{\sqrt{2}} \; 
\end{equation}
$\eta_0$ is the unperturbed eigenstate corresponding to $n=0$.  There
are no decoupled eigenstates.

The exact masses $\lambda$ of the coupled states are determined by
\eq{eigenvalues}, which in our case becomes
\begin{equation}
\label{majoval}
\left[ \lambda^2 (1-x) -A \right](1-x) = \MM^2 x^2 \; ,
\end{equation}
where 
\begin{equation}
\label{notation}
x = \pi R |\mbb|^2 \frac{\cot \pi
R\sqrt{\lambda^2-\MM^2}}{\sqrt{\lambda^2-\MM^2}} \; .
\end{equation}
By solving for $x$, \Eq{majoval} can be written as
\begin{equation}
\label{majoval2}
\pi R |\mbb|^2 \frac{\cot \pi
R\sqrt{\lambda^2-\MM^2}}{\sqrt{\lambda^2-\MM^2}} =
\frac{\lambda}{\lambda\pm \MM} \; .
\end{equation}

The corresponding mass eigenstates are
\begin{equation}
\label{majostate}
\hat\nu = \frac{1}{\sqrt{N}} \left[ \nu + 
\sum_{n\in\Z} \frac{\bar
\mbb}{\lambda^2-\MM^2-n^2/R^2}\left(i\frac{n}{R}\eta_n +
\frac{\MM}{1-x}\xi_n\right)\right] \; ,
\end{equation}
where $N$ is a normalization factor.

% \bibliographystyle{utcaps}
% \bibliography{abbrev,biblio,hep,tmp,pro}

\providecommand{\href}[2]{#2}\begingroup\raggedright

\end{document}